\documentstyle[12pt]{article}
\begin{document}
\thispagestyle{empty}
\begin{flushright}
MPI-Ph/93-56\\
July 1993
\end{flushright}
\bigskip\bigskip\begin{center}
{\bf \Huge{Functional Integral Approach to the N-Flavor Schwinger Model}}
\end{center}  \vskip 1.0truecm
\centerline{\bf
Christof Gattringer${}^*$ and Erhard Seiler}
\vskip5mm
\centerline{Max-Planck-Institut f\"{u}r
 Physik, Werner-Heisenberg-Institut}
\centerline{F\"ohringer Ring 6, 80805 Munich, Germany}
\vskip 2cm
\bigskip \nopagebreak \begin{abstract}
\noindent
We study massless $QED_2$ with $N$ flavors using path integrals.
We identify the sector
that is generated by the $N^2$ classically conserved vector currents.
One of them (the $U(1)$ current) creates a massive particle,
while the others create massless ones. We show that the mass spectrum obeys a
Witten-Veneziano type formula.
Two theorems on $n$-point functions
clarify the structure of the Hilbert space. Evaluation of the
Fredenhagen-Marcu order parameter indicates that a
confining force exists only
between charges that are integer multiples of $\pm Ne$, whereas
charges that are nonzero $\mbox{mod}(N)$ screen their confining forces and lead
to non-vacuum sectors. Finally we identify operators that violate clustering,
and decompose the theory into clustering $\theta$ vacua.
\end{abstract}
\vskip 3cm
\bigskip \nopagebreak \begin{flushleft} \rule{2 in}{0.03cm}
\\ {\footnotesize \ ${}^*$ Also at Institut f. Theoretische Physik d.
Universit\"at Graz, Austria }
\end{flushleft}

\newpage\setcounter{page}1

\section{Introduction}
The Schwinger model is a laboratory to study many important aspects of more
realistic models of particle physics such as QCD.
It shows confinement, mass generation of its would-be Goldstone particle
via the axial anomaly -- thereby solving its $U(1)$ problem,
it also has topological sectors and a vacuum angle $\theta$ dual to them.
While these features are well known and understood in the case of Schwinger's
original one-flavor model, in a model with several flavors the situtation
is much less trivial and has not been investigated in much detail so far. This
$N$-flavor Schwinger model has in fact more features in common with four
dimensional QCD: On the classical level it has a symmetry group
$U(N)_L\times U(N)_R$ that is broken down by the anomaly to
$SU(N)_L\times SU(N)_R\times U(1)_V$ just like in QCD.

In this paper we study this model using the euclidean functional integral.
The boundary conditions we use (vector potentials decaying at infinity
sufficiently fast to be square integrable)
correspond to zero topological charge for the gauge field. This leads to
an infinite volume vacuum state enjoying chiral invariance, but violating the
cluster decomposition property. It turns out that by explicit construction
this vacuum state can be decomposed into a mixture of clustering states
labeled by an angle $\theta$, the so-called vacuum angle, which is
conventionally
introduced as a dual variable to the topological charge (first Chern number)
of the gauge field.

We compute the correlation functions of the vector/pseudovector currents
as well as the
scalar/pseudoscalar densities and find that the theory describes a multiplet
of one massive and $N^2-1$ massless pseudoscalar `mesons'. There is a relation
between the masses that looks exactly like the so-called Witten-Veneziano
formula for the pseudoscalar meson masses in QCD; in fact it allows to
give a precise meaning to this formula with proper attention to the short
distance fluctuations of the topological density that play an essential role.

The model also reveals an interesting superselection structure.
Computation of the so-called Fredenhagen-Marcu confinement parameter suggests
that there are superselection sectors labeled by a charge that is defined as
an integer
$\mbox{mod}(N)$. This means that the confining Coulomb potential
does not prevent
operators of fermion number $\neq 0 \; \mbox{mod}(N)$ from creating new charged
sectors; their electric charge seems to be `screened'. Of course QCD
is not expected to share this property.

The model is still trivial in the sense that there is no real interaction
between its particles, but it is not Gaussian like the one-flavor model.
Its vacuum sector turns out to be a tensor product of the Fock space of a free
massive scalar particle with the representation space of the
$SU(N)_L \times SU(N)_R$ current (Kac-Moody) algebra induced by $N$ massless
Dirac fermions.

We would like to make a remark about the level of rigor of this paper:
Our aim is to elucidate the physical content by explicit calculation.
Therefore we do not spend time to justify all the manipulations involved
in deriving the results with full mathematical rigor, even though it would
not be too difficult to do so. In \cite {seiler} it is sketched how this is
to be done for the one-flavor model; the $N-$flavor models does not pose
any fundamentally new problems in this respect.

There are some open problems left: For instance it should be expected that
the subspace created form the vacuum by the scalar/pseudovector currents is
identical to the one created by the vector/pseudoscalar densities;
we did not attempt to prove that. Likewise the
structure of superselection sectors suggested by the Fredenhagen-Marcu
parameter has not been established rigorously, because that parameter
by its nature has only suggestive value.

The paper is organized as follows:
In the next section we define the model and introduce our notations.
Furthermore we derive some basic expressions for
generating functionals and Green's functions.
In Section 3 we investigate the sector that is generated by vector currents.
The mass spectrum that we obtain fits a Witten-Veneziano type formula,
which we discuss in Section 4.
To get more insight into the subspace generated by the vector currents
we derive general expressions for their $n$-point functions in section 5.
The next section is dedicated to
the problem of confinement. We evaluate the Fredenhagen-Marcu order parameter
for a $e^+ e^-$ system and higher products of fermion fields. Finally we
decompose the model into clustering theories and thereby obtain the
vacuum angle.

\section{Setup}
We use the following representation of the 2-d Euclidean $\gamma$-matrices.
\begin{equation}
\gamma_1 =  \left( \begin{array}{cc}
0 & 1 \\ 1 & 0 \end{array} \right) \; \; , \; \;
\gamma_2 =  \left( \begin{array}{cc}
0 & -i \\ i & 0 \end{array} \right) \; \; , \; \;
\gamma_5 = i \gamma_2 \gamma_1 =  \left( \begin{array}{cc}
1 & 0 \\ 0 & -1 \end{array} \right)
\end{equation}
They obey the commutation relations $\{\gamma_\mu , \gamma_\nu \} =
2 \delta_{\mu \nu} \: ; \: \mu,\nu = 1,2,5 $.
The Green's function $G(x,y;A)$ for fermions in a background field $A$
obeys the equation
\begin{equation}
\gamma_\mu ( \partial_\mu - i e A_\mu ) G(x,y;A) = \delta(x,y) \; .
\end{equation}
In two dimensions it can be solved explicitely  \cite{schwinger},
provided the background field satisfies some mild regularity and
falloff conditions; the solution is
\begin{equation}
G(x,y;A) = e^{ie[ \Phi(x) - \Phi(y) ]} \;  G^0 (x-y)
\end{equation}
where
\begin{equation}
\Phi(x) = - \int d^2\!z D(x-z) \Big[ \partial_\mu A_\mu (z)
+ i \gamma_5 \varepsilon_{\mu \nu} \partial_\mu A_\nu (z) \Big] \; .
\end{equation}
$G^0(x)$ is the Green's function for zero background field
\begin{equation}
G^0(x) = \frac{1}{2\pi} \frac{\gamma_\mu x_\mu}{x^2} \; .
\end{equation}
$D(x)$ denotes the Green's function for free massless bosons
($-\triangle D(x) = \delta(x)$).

The action for the $N$-flavor Schwinger model reads
\begin{equation}
S[\psi,\overline{\psi},A] = S_g [A] + S_f [ \psi,\overline{\psi},A] \; ,
\end{equation}
with
\begin{equation}
S_g [A] = \int d^2x \left[ \frac{1}{4} F_{\mu \nu}(x) F_{\mu \nu}(x)
+ \frac{1}{2} \lambda \left( \partial_\mu A_\mu(x) \right)^2 \right] =:
\frac{1}{2} ( A, Q A).
\end{equation}
The kernel $Q$ of the quadratic form we introduced is given by \\
$Q = (1 - \lambda) \partial \otimes \partial - \triangle$ .
We work in Landau gauge, which means $\lambda \rightarrow \infty$
after the integration over the gauge field.

The fermionic part of the action is a generalization of the one flavor
model
\begin{equation}
S_f[ \psi, \overline{\psi},A] = \sum_{a=1}^N \int d^2x
\overline{\psi}^{(a)}(x) \gamma_\mu \Big[ \partial_\mu - ie A_\mu(x) \Big]
\psi^{(a)} \;  .
\end{equation}
$\overline{\psi}^{(a)}, \psi^{(a)}$ are 2-component spinors with a flavor
index $a = 1,2,...N$.
It is useful to introduce sources for the fermions and to define the generating
functional
\begin{equation}
Z[\eta,\overline{\eta}] := \int [d\psi][d\overline{\psi}][dA]
e^{-S[\psi,\overline{\psi},A] + \sum_a \Big\{
( \overline{\eta}^{(a)},\psi^{(a)} )
+ (\overline{\psi}^{(a)},\eta^{(a)} ) \Big\} }  \; .
\end{equation}
Vacuum expectation values can be obtained as functional derivatives
with respect to $\eta , \overline{\eta}$. Performing formally the Berezin
integral over the Grassmann variables leads to
\begin{equation}
Z[\eta,\overline{\eta}] = \int[dA] e^{-S_g[A]}  \mbox{det}[ \not\!{\partial}
- ie \not{\!\!A} ]^N  e^{\sum_a \int d^2x d^2y \: \overline{\eta}^{(a)}\!(x)
\: G(x,y;A) \: \eta^{(a)}\!(y) }\; .
\end{equation}
Note that the formal expression $\mbox{det}[ \not\!{\partial}
- ie \not{\!\!A} ]$ is only defined when an ultraviolet and infrared
cutoff (for instance a finite space-time lattice)
is introduced. The determinant can then be normalized to 1 for $e = 0$, by
replacing it with $\mbox{det}[1 - eK(A)]$ where
$K(A) = i \not{\!\!A} {\not\!{\partial}}^{-1}$.
In two dimensions this determinant can be computed explicitely
\cite{schwinger}, using the idea of regularized fermion determinants
(for a review see e.g. \cite{simon}, for the $\mbox{QED}_2$ case
\cite{seiler}).
If we assume that the vector potential $A$ satisfies
some mild regularity and falloff conditions at infinity to make it
square integrable, the answer is
\begin{equation}
\mbox{det}[ 1 - eK(A) ] = e^{-\frac{e^2}{2\pi}
\parallel A^T \parallel_2^2 } \; ,
\end{equation}
where $A_\mu^T$ is the transverse part of the gauge field
\begin{equation}
A_\mu^T = A_\mu - \frac{\partial_\mu \partial_\nu}{\triangle} A_\nu \; .
\end{equation}
We combine the logarithm of the
determinant and the gauge field action in one common quadratic
form for $A_\mu$. This gives rise to
a Gaussian measure $d\mu_C [A] = \frac{1}{Z} [dA]
\exp( -\frac{1}{2} ( A, C^{-1} A ) ) $ for the gauge field
with covariance $C$ defined by
\begin{equation}
( A, C^{-1} A ) := ( A, Q A ) + e^2 \frac{N}{\pi}
\parallel\!A^T\!\parallel^2_2 \; .
\end{equation}
A few lines of algebra lead to
\begin{equation}
C^{-1}_{\mu \nu} = \Big( -\triangle + e^2 \frac{N}{\pi} \Big)
\Big( \delta_{\mu \nu} - \frac{\partial_\mu \partial_\nu}{\triangle} \Big) -
\partial_\mu \partial_\nu  \lambda  \; .
\end{equation}
Inverting and taking the limit $\lambda \rightarrow \infty$ gives
\begin{equation}
C_{\mu \nu} = \frac{1}{-\triangle + e^2 \frac{N}{\pi}} \Big( \delta_{\mu \nu}
- \frac{\partial_\mu \partial_\nu}{\triangle} \Big) \;  .
\end{equation}
Besides this form of the measure for the gauge field, we compute
the measure for a field $\varphi$ which is a linear functional of the
gauge field. This leads to a very simple
dependence of the propagator on the gauge field.
$G(x,y;A)$ depends on $\Phi[A]$ via the exponential
$\exp(ie[\Phi(x) - \Phi(y) ] )$ (see equation (3)).
$\Phi$ in general depends on the gauge field
through the terms $\partial_\mu A_\mu$
and $\varepsilon_{\mu \nu} \partial_\mu A_\nu$, but
the first term vanishes (with probability 1) in Landau gauge so
that $\Phi$ becomes
a function of $\varepsilon_{\mu\nu}\partial_\mu A_\nu$ alone.
To get the spinor structure
of the propagator explicitly,
we define (after dropping the $\partial_\mu A_\mu$ term in $\Phi$ )
\begin{equation}
\Phi(x) =: i \gamma_5 \varphi(x) \; ,
\end{equation}
where
\begin{equation}
\varphi(x) = - \int d^2z D(x-z) \varepsilon_{\mu \nu} \partial_\mu A_\nu(z)\; .
\end{equation}
This implies
\begin{equation}
\varphi = \frac{\varepsilon_{\mu \nu} \partial_\mu }{\triangle} A_\nu \; .
\end{equation}
Using this equation one obtains the Gaussian measure
$d\mu_{\tilde{C}}[\varphi]$ for the field $\varphi$,
induced by the measure for the gauge field with
covariance $\tilde{C}$ given by
\begin{equation}
\tilde{C} = \frac{-1}{ ( -\triangle + N \frac{e^2}{\pi} ) \triangle} \; .
\end{equation}
Expressed in terms of $\varphi$ the propagator reads
\[
G(x,y;\varphi) = e^{-e \gamma_5 [ \varphi(x) - \varphi(y) ] } \; G^0 (x-y) = \]
\begin{equation}
\frac{1}{2\pi}\frac{1}{(x-y)^2} \left( \begin{array}{cc}
0 & e^{-e[\varphi(x)-\varphi(y)]} \; \overline{(\tilde{x}-\tilde{y})} \\
e^{+e[\varphi(x)-\varphi(y)]} \; (\tilde{x}-\tilde{y}) & 0
\end{array} \right) \; ,
\end{equation}
where we introduced the complex coordinate
\begin{equation}
\tilde{x}= x_1 + i x_2 \; .
\end{equation}
\section{Vector currents}
In this section we derive the 2-point functions of the vector/pseudovector
currents and obtain thereby information about the particle spectrum of the
theory. Note that the axial vector currents are simply linear combinations
of the vector currents due to the fact that in two dimensions
$\gamma_5\gamma_{\mu\nu}= i \varepsilon_{\mu\nu}\gamma_\nu$.

\subsection{Bosonizing a Cartan subalgebra of $U(N)$}
It was noticed in \cite{belvedere} that the $U(1)$ current together
with the other elements of a Cartan subalgebra of $U(N)_{flavor}$ can be
bosonized in terms of free scalar fields. This result is the first step
in our analysis of the particle structure of the model and therefore we
reproduce this result in terms of the functional integral formalism.

Define vector currents
\begin{equation}
j^{(b)}_\mu(x) := \overline{\psi}^{(b)}(x) \gamma_\mu \psi^{(b)}(x) \;
, b = 1,2,...N \; .
\end{equation}
In principle one would have to regularize these currents by e.g.
a point splitting prescription, but we can use the fact that the
fermion current couples to an external field $a_{\mu}$ the same way
as to the quantized field $A_{\mu}$ and thereby avoid all
problems due to short distance singularities.
We couple the currents to source fields $a^{(b)}_\mu(x)$ obeying
$\partial_\mu a^{(b)}_\mu = 0$, and evaluate the characteristic
functional using the explicit form of the determinant
\[
\Big\langle e^{ie \sum_{b=1}^N (j^{(b)}_\mu,a^{(b)}_\mu)}\Big\rangle =
\frac{1}{Z}\int[dA]e^{-S_g[A]} \prod_{b=1}^N
\mbox{det}\Big[ 1 - eK( A + a^{(b)}) \Big] =
\]
\[
e^{-\frac{e^2}{2\pi}\sum_b \parallel a^{(b)} \parallel^2_2}
\int d\mu_C[A] e^{-(A,\Lambda)} =
e^{-\frac{e^2}{2\pi}\sum_b \parallel a^{(b)} \parallel^2_2} \;\;
e^{\frac{1}{2}(\Lambda,C\Lambda)} \; ,
\]
\begin{equation}
\Lambda_\mu = \frac{e^2}{\pi}\sum_{b=1}^{N} a^{(b)}_\mu \; .
\end{equation}
After some reordering of the terms in the last equation one
ends up with the following expression for the characteristic functional
\begin{equation}
\Big\langle e^{-ie\sum_b(j^{(b)}_\mu,a^{(b)}_\mu)} \Big\rangle =
e^{-\frac{e^2}{2\pi}\sum_{b,c}
(\varepsilon_{\mu \nu} \partial_\mu a^{(b)}_\nu , M^{(b,c)}
\varepsilon_{\rho \sigma} \partial_\rho a^{(c)}_\sigma ) } \; ,
\end{equation}
where $M$ is given by
(using matrix notation in flavor space)
\[
M = \frac{-e^2/\pi}{(-\triangle + e^2\frac{N}{\pi}) \triangle} R
+ \frac{1}{-\triangle + e^2\frac{N}{\pi}}{\bf 1} \; , \]
\begin{equation}
R = \left( \begin{array}{ccccc}
N-1 & -1 & . & . & -1 \\
-1 & N-1 & -1 &   & .  \\
 . & -1  & . & . & .  \\
 . &     & . & . & -1 \\
-1 & .   & . & -1 & N-1
\end{array} \right) \; .
\end{equation}
The numerical matrix $R$ can be diagonalized by an orthogonal matrix
$U$ given by $U^T= ( r^{(1)},r^{(2)}, ..... r^{(N)})$ where the
$r^{(l)}$ are the normalized eigenvectors of $R$.
The corresponding eigenvalues are $0,N,N,.....N$, implying
$U R U^T = \mbox{diag}(0,N,....N)$.
This allows to diagonalize the covariance $M$. Define
\begin{equation}
K := U M U^T = \left( \begin{array}{ccccc}
\frac{1}{-\triangle +e^2\frac{N}{\pi}} & & & & \\
 & \frac{1}{-\triangle} & & & \\
 & & . & & \\
 & & & . & \\
 & & & & \frac{1}{-\triangle} \end{array} \right) \; .
\end{equation}
We now choose the sources $a_\mu^{(b)}$ in flavor space proportional to
one of the eigenvectors $r^{(l)}$.
This corresponds to a change of the basis in flavor space
and allows to express expectation values
of the new vector currents in terms of free bosons. Define
\begin{equation}
J^{(l)}_\mu := \overline{\psi} \gamma_\mu H^{(l)} \psi \; ,
\end{equation}
where the $N\times N$ matrices $H^{(l)}$
are generators of a Cartan subalgebra of $U(N)_{flavor}$
\[
H^{(1)} = \frac{1}{\sqrt{N}} \; {\bf 1} \;  , \] \[
H^{(2)} = \frac{1}{\sqrt{N-1+(N-1)^2}} \; \mbox{diag}
(1,1, .... 1, -N+1 ) \; , \] \[
H^{(3)} = \frac{1}{\sqrt{N-2+(N-2)^2}} \; \mbox{diag}
(1,1, .....1, -N+2 , 0) \; , ...... \]
\begin{equation}
H^{(N)} = \frac{1}{\sqrt{2}} \; \mbox{diag}
(1,-1,0,0, ...... 0) \; .
\end{equation}
The diagonals of these generators are the eigenvectors of $M$.
Using equation (24) one easily finds the generating functional for the
$J^{(l)}_\mu$
\[
\Big\langle e^{ie \left( J^{(l)}_\mu , b^{(l)}_\mu \right) }\Big\rangle =
e^{-\frac{e^2}{2\pi} \left( r^{(l)} \varepsilon_{\mu \nu}
\partial_\mu b_\nu^{(l)}, M  r^{(l)} \varepsilon_{\rho \sigma}
\partial_\rho b_\sigma^{(l)} \right) } = \]
\begin{equation}
e^{-\frac{e^2}{2\pi} \left( \varepsilon_{\mu \nu} \partial_\mu b_\nu^{(l)},
K_{l l} \, \varepsilon_{\rho \sigma} \partial_\rho b_\sigma^{(l)} \right) }
= \Big\langle e^{ ie \frac{1}{\sqrt{\pi}} \left( \varepsilon_{\mu \nu}
\partial_\nu \varphi^{(l)} , b_\mu^{(l)} \right) }\Big\rangle_B \; ,
\end{equation}
where $b_\mu^{(l)}$ is a source for the currents $J_\mu^{(l)}$.
The index $l$ now labels the `meson' fields (27).
$\langle \cdot \rangle_B$ denotes the Gaussian expectation values of
an $N$-tuple of scalar fields $\varphi^{(l)} , l = 1,2,...N$ with covariance
$K$ given by equation (26).
Hence we can identify
\begin{equation}
J^{(l)}_\mu = \frac{1}{\sqrt{\pi}}\varepsilon_{\mu \nu}
\partial_\nu \varphi^{(l)} \; .
\end{equation}
The last equation is the
bosonization formula of a Cartan subalgebra of $U(N)$ \cite{belvedere}.
\subsection{More meson fields}
In analogy to the construction of meson states in QCD, one
can define vector currents for all the generators of $U(N)$.
A convenient basis of the Lie algebra of $U(N)$ is given by the
$N(N-1)/2$ generators $H^{(l)}$ of the form
\begin{equation}
\frac{1}{\sqrt{2}} \left( \begin{array}{cccccc}
 & & & & & \\
 & & & &1& \\
 & & & & & \\
 &1& & & & \\
 & & & & & \\
 & & & & &
\end{array} \right) \; ,
\end{equation}
and $N(N-1)/2$ of the form
\begin{equation}
\frac{1}{\sqrt{2}} \left( \begin{array}{cccccc}
 & & & & & \\
 & & & &-i& \\
 & & & & & \\
 &i& & & & \\
 & & & & & \\
 & & & & &
\end{array} \right) \; .
\end{equation}
The corresponding vector currents are
\begin{equation}
J_\mu^{(l)}(x) := \overline{\psi}(x)H^{(l)}\psi(x) .
\end{equation}
Again no point splitting has to be taken into account, since
only different flavors (which cannot contract) sit at one space-time point
and therefore there is no short distance singularity.

First we notice that all $N^2$ vector currents generate orthogonal
states
\begin{equation}
\Big\langle J_\mu^{(l)}(x) J_\nu^{(l^\prime)}(y)\Big\rangle =
\delta_{l l^\prime} \;\; {\cal F}_{\mu \nu}^{(l)} (x,y)
\;\;\;\; l = 1,2,...N^2 \; ,
\end{equation}
where $\cal F$ is the two point function. For the Cartan currents
(34) follows dircectly from the bosonization. To prove it for the set of all
$N^2$ currents one has to take functional derivatives of
the generating functional (10)
with respect to the fermion sources $\eta, \bar{\eta}$.
If $l \neq l^\prime$, either the different flavors do not contract
entirely, or terms with opposite sign cancel.

In the case $ l = l^\prime , l = N+1,N+2,...N^2$,
functional derivation leads to
(take e.g. $\mu = \nu = 1$)
\begin{equation}
{\cal F}_{1 1} = - \int d\mu_{\tilde{C}}[\varphi] \Big\{
G_{21}(x,y;\varphi)G_{21}(y,x;\varphi) +
G_{12}(x,y;\varphi)G_{12}(y,x;\varphi) \Big\} \; .
\end{equation}
Using the explicit form (20) for $G$, one immediately sees that
the exponentials involving $\varphi$ cancel. Integration over
$\varphi$ simply gives a factor 1. The same is true for arbitrary
$\mu,\nu$. Hence the two point function is the same as for free, massless
fermions. It can be expressed in terms of derivatives of the
propagator of a free massless boson, giving rise to the same expression as
was obtained for the Cartan subalgebra, i.e. for $l = 2,3,...N$.
Putting things together
we conclude for the two-point functions of the vector currents
\begin{equation}
\Big\langle J_\mu^{(l)}(x) J_\nu^{(l^\prime)}(y) \Big\rangle =
\delta_{l l^\prime} \;\; \frac{1}{\pi} \varepsilon_{\mu \rho} \partial x_\rho
\,\varepsilon_{\nu \sigma} \partial y_\sigma
\;\Big\langle \varphi^{(l)}(x) \varphi^{(l)}(y) \Big\rangle_B \; ,
\end{equation}
where the scalar field $\varphi^{(1)}$ has mass $m = e\sqrt{\frac{N}{\pi}}$
and $\varphi^{(l)}, l=2,3,...N^2$ are massless. In addition we obtained
explicit expressions for arbitrary $n$-point functions of the
Cartan currents $(l=1,2,...N)$ via the bosonization prescription (30).
In Section 5 we will obtain the $n$-point functions for all the currents.
\section{Mass spectrum and Witten-Veneziano type formulas}

In 1979 Witten \cite{witten} proposed a formula that relates the mass of the
$\eta^\prime$ meson to the topological susceptibility of quarkless QCD.
The purpose of this was to show how QCD solves its $U(1)$ problem,
without having to
bring instantons into the discussion. The derivation given by Witten
has some problems (terms of equal sign are cancelled against each other)
as was pointed out by Seiler and Stamatescu \cite{seilerstam}.
But it was also noted there that for the one flavor Schwinger model
the formula is nevertheless true, if the topological susceptibility is
interpreted apppropriately, with proper attention to contact terms
(the fact that therefore the correct treatment of short distance fluctuations
of the topological charge density is essential makes of course the
application to lattice gauge theories cumbersome, to say the least).

The $N$ flavor Schwinger model is a convenient laboratory to study the $U(1)$
problem in a more general setting, much more reminiscent of QCD, because
it has both flavor changing axial currents that are conserved
and a flavor non-changing one whose conservation is violated by the
axial anomaly.
The mass spectrum we obtained in the last section contains both massless
Goldstone particles related to the conserved axial currents and
a massive particle related to the anomalous $U(1)$ axial current.
So it has the features
expected to be present in the meson sector of QCD in the chiral limit.
Hence it is natural to check if our model satisfies a formula of the
Witten-Veneziano type.

Here we cite the more general formulation of Veneziano
(\cite{veneziano},\cite{smit})
which reduces to Witten's form when setting all meson masses (except
$m_{\eta^\prime}$) equal zero. It is given by
\begin{equation}
m^2_{\eta^\prime} - \frac{1}{2} m^2_{\eta} - \frac{1}{2} m^2_{\pi^0} =
\frac{12}{f^2_\pi} \chi \;.
\end{equation}
$f_\pi$ denotes the meson decay constant. $\chi$ has a more
subtle meaning. The interpretation as topological susceptibility
faces some problems, as was pointed out by Seiler and Stamatescu
\cite{seilerstam}. But a formula of this type can be derived if
$\chi$ is interpreted as the contact term that occurs in the spectral
representation of the two point function of the topological charge density
(compare also the appendix in \cite{witten}, and the derivation given
by Smit and Vink \cite{smit}). Here we use the form that was obtained
by Seiler and Stamatescu for an arbitrary number of flavors $N$ and
vanishing quark masses, which implies $m_\eta = m_{\pi^0} = 0$. It reads
\begin{equation}
m^2_{\eta^\prime} = \frac{4 N}{\mid f_{\eta^\prime} \mid^2} P(0) \; .
\end{equation}
$P(0)$ denotes the announced contact term. (38) contains the $\eta^\prime$
decay constant $f_{\eta^\prime}$, which shows up also in Witten's
derivation, and is replaced later by $f_\pi$ to which it is approximately
equal.

In $\mbox{QED}_2$ the topological charge density $q(x)$ is given by
\begin{equation}
q(x) = \frac{e}{2\pi} F_{12}(x) \; .
\end{equation}
The topological susceptibility is defined to be
\begin{equation}
\chi_{top} = \int \langle q(x) q(0) \rangle dx = \frac{e^2}{(2\pi)^2}
\int \langle F_{12}(x) F_{12}(0) \rangle dx =
\frac{e^2}{2\pi} \hat{G}_{FF}(0) \; ,
\end{equation}
where $G_{FF}$ denotes the $F_{12}$ propagator, and $\hat{G}_{FF}(0)$
it's Fourier transform at zero momentum. Since
\begin{equation}
F_{12}(x) = \varepsilon_{\mu \nu} \partial_\mu A_\nu(x) \; ,
\end{equation}
the propagator $G_{FF}$ is given by
\begin{equation}
G_{FF} = - \varepsilon_{\mu \nu} \partial_\nu
C_{\mu \rho} \varepsilon_{\rho \sigma} \partial_\sigma \; ,
\end{equation}
where $C$ is the gauge field propagator (15).
Inserting $C$ and transforming to momentum space we obtain
\begin{equation}
\hat{G}_{FF}(p) = \frac{p^2}{p^2 + e^2\frac{N}{\pi}} =
1 - \frac{e^2\frac{N}{\pi}}{p^2 + e^2\frac{N}{\pi}} =
1 - \int_0^\infty \frac{d\rho(\mu^2)}{p^2 + \mu^2} \; .
\end{equation}
In the last step we made the spectral integral explicit. One nicely
sees that the spectral measure
\begin{equation}
d\rho(\mu^2) = \delta(\mu^2 - m^2) d\mu^2
\end{equation}
is `dominated' by the contribution of the $\eta^\prime$, which should
of course
be identified with our massive particle. Hence we have
\begin{equation}
m^2_{\eta^\prime} = m^2  = e^2\frac{N}{\pi} \; .
\end{equation}
{}From (43) one immediately reads
off the contact term $P(0)$ in the spectral decomposition of
$\chi_{top}$
\begin{equation}
P(0) = \frac{e^2}{4\pi^2} \; .
\end{equation}
The last missing ingredient is the decay constant $f_{\eta^\prime}$.
It is defined by
\begin{equation}
f_{\eta^\prime} = m^{-2} \langle 0 \mid
\partial_\mu J^{(1)} \mid \eta^\prime \rangle \; .
\end{equation}
The anomaly equation for the current
$j^{(a)}_\mu = \overline{\psi}^{(a)} \gamma_\mu \psi^{(a)}$ reads
(see eg. \cite{seiler} )
\begin{equation}
\partial_\mu j^{(a)}_\mu = 2 q \; .
\end{equation}
Using the definition of the $U(1)$ current (27),(28) one obtains the anomaly
equation\footnote{Note that in our definition of the $U(1)$ current there is
an extra factor $1/\sqrt{N}$ which modifies (47),(49) by this factor, compared
to the usual notation.}
\begin{equation}
\partial_\mu J^{(1)} = 2 \sqrt{N} q \; ,
\end{equation}
which we insert in (47) to end up with
\begin{equation}
f_{\eta^\prime} =
m^{-2} \; 2 \sqrt{N} \langle 0 \mid q \mid \eta^\prime \rangle =
m^{-2} \; 2 \sqrt{N} \langle 0 \mid q \; \frac{i}{m} F_{12} \mid 0 \rangle \; .
\end{equation}
In the last step we generated the $| \eta^\prime \rangle$ state as
$Z^{\frac{1}{2}} F_{12} | 0 \rangle$ with the normalization condition
\begin{equation}
\hat{G}_{\eta^\prime \eta^\prime} \stackrel{!}{=} \frac{1}{p^2 + m^2}
+ \mbox{contact term} \; ,
\end{equation}
giving rise to $Z = -\frac{1}{m^2}$.
We end up with
\begin{equation}
f_{\eta^\prime} = i \frac{1}{\sqrt{\pi}} \; .
\end{equation}
Insertion in (38) of the quantities  obtained gives an identity.
This explicit computation shows that eq. (38) holds in massless
$\mbox{QED}_2$ with $N$ flavors. But one can also interprete the result
as a verification of the original form of the Witten-Veneziano formula,
because the topological susceptibility of the quenched theory reduces
to the contact term. It is not true, however, that the topological
susceptibility appearing in the formula expresses only a property of the long
distance fluctuations of the topological density.
\section{N-point functions}
To gain more insight into the sector generated by the vector currents
we compute explicit expressions for the connected $n$-point
functions $C_n$. Considering fully connected correlations
has the advantage of avoiding contractions of fermions
at the same point. Therefore we can avoid using test functions as
would be required in principle by the distributional nature of the fields.
Let
\[
C_{n+1} := \Big\langle J_{\mu_0}^{(l_0)}(x_0) J_{\mu_1}^{(l_1)}(x_1) .....
 J_{\mu_n}^{(l_n)}(x_n) \Big\rangle_c =  \]
\[ \sum_{a_i,b_i} H^{(l_0)}_{a_0 b_0} H^{(l_1)}_{a_1 b_1} .....
H^{(l_n)}_{a_n b_n} \sum_{\alpha_i,\beta_i}
(\gamma_{\mu_0})_{\alpha_0 \beta_0} (\gamma_{\mu_1})_{\alpha_1 \beta_1} .....
(\gamma_{\mu_n})_{\alpha_n \beta_n} \]
\begin{equation} \Big\langle
\overline{\psi}^{(a_0)}_{\alpha_0}(x_0) \psi^{(b_0)}_{\beta_0}(x_0)
\overline{\psi}^{(a_1)}_{\alpha_1}(x_1) \psi^{(b_1)}_{\beta_1}(x_1) .....
\overline{\psi}^{(a_n)}_{\alpha_n}(x_n) \psi^{(b_n)}_{\beta_n}(x_n)
\Big\rangle_c \; .
\end{equation}
Nonvanishing contributions occur only if the color indices can form a
closed chain (e.g. $b_0 = a_1,b_1 = a_2, ... b_{n-1} = a_n,b_n = a_0$). The
corresponding factor is simply the trace over the flavor matrices $H^{(l)}$.
To find all possible contributions one has to sum over all permutations $\pi$,
keeping the first term fixed.
\[
C_{n+1} =  -\!\!\sum_{\pi(1,2,...n)}\!\mbox{Tr} \Big[
H^{(l_0)} H^{(l_{\pi(1)})} ..... H^{(l_{\pi(n)})} \Big]
\sum_{\alpha_i,\beta_i}
(\gamma_{\mu_0})_{\alpha_0 \beta_0} (\gamma_{\mu_1})_{\alpha_1 \beta_1} .....
(\gamma_{\mu_n})_{\alpha_n \beta_n} \]
\begin{equation}
\int\!d\mu_{\tilde{C}}[\varphi]
G_{\beta_0 \alpha_{\pi(1)}} (x_0,x_{\pi(1)};\varphi)
G_{\beta_{\pi(1)} \alpha_{\pi(2)}} (x_{\pi(1)},x_{\pi(2)};\varphi) .....
G_{\beta_{\pi(n)} \alpha_0} (x_{\pi(n)},x_0;\varphi)
\end{equation}
Since in the chosen representation the $\gamma$ matrices and the propagator
$G(x,y;\varphi)$ have only off-diagonal entries (cf. equation(20)),
we find the following chain of implications for e.g. $\beta_0 = 1$
\begin{equation}
\beta_0 = 1 \Rightarrow \alpha_{\pi(1)} = 2 \Rightarrow \beta_{\pi(1)} = 1
\Rightarrow \alpha_{\pi(2)} =2 ..... \Rightarrow \beta_{\pi(n)} = 1
\Rightarrow \alpha_0 = 2 .
\end{equation}
When starting with $\beta_0 = 2$ one ends up with the opposite result of all
$\beta_i = 2$ and all $\alpha_i = 1$. Besides those two no other
nonvanishing terms contribute.
Again one finds that all dependence on $\varphi$ cancels, and
only free propagators $G^0$ remain. Using $G^0(x) = \frac{1}{2\pi}
\frac{\gamma_\mu x_\mu}{x^2}$ and making use of the complex notation
(21) we obtain
\[
C_{n+1} = -\frac{1}{(2\pi)^{n+1}}
\sum_{\pi(1,2...n)} \mbox{Tr}\Big[H^{(l_0)} H^{(l_{\pi(1)})} .....
H^{(l_{\pi(n)})}\Big]
\]
\begin{equation}
\left\{ \prod_{i=0}^n (\gamma_{\mu_i})_{21}
\frac{1}{\tilde{x}_0 - \tilde{x}_{\pi(1)}}
\frac{1}{\tilde{x}_{\pi(1)} - \tilde{x}_{\pi(2)}} .....
\frac{1}{\tilde{x}_{\pi(n)} - \tilde{x}_0} \; + \; c.c. \right\} \; ,
\end{equation}
where $c.c.$ denotes complex conjugation.

Using this basic formula, we construct for $N\geq2$ and arbitrary $n$
fully connected
$n$-point functions that do not vanish; in other words we show that the
currents are not Gaussian. For simplicity we give the
construction for the case of $N=2$. Since for arbitray $N > 2$
generators with the same commutation relations exist, it is obvious how
to generalize the construction to general $N$.
The generators we are using are the Pauli matrices
up to a normalization factor.
To distinguish them from arbitrary generators $H^{(l)}$ we denote the
special set of matrices needed in the construction
\begin{equation}
\tau^{(1)} := \frac{1}{\sqrt{2}} \left( \begin{array}{cc}
0 & 1 \\
1 & 0 \end{array} \right) \; , \;
\tau^{(2)} := \frac{1}{\sqrt{2}} \left( \begin{array}{cc}
0 & -i \\
i & 0 \end{array} \right) \; , \;
\tau^{(3)} := \frac{1}{\sqrt{2}} \left( \begin{array}{cc}
1 & 0 \\
0 & -1 \end{array} \right) \;.
\end{equation}
Define
\[
F_{n+1}(\tilde{x}_0,\tilde{x}_1,... \tilde{x}_n) :=
\]
\begin{equation}
\sum_{\pi(1,2,...n)} \mbox{Tr}\Big[\tau^{(l_0)} \tau^{(l_{\pi(1)})} .....
\tau^{(l_{\pi(n)})} \Big]
\frac{1}{\tilde{x}_0 - \tilde{x}_{\pi(1)}}
\frac{1}{\tilde{x}_{\pi(1)} - \tilde{x}_{\pi(2)}} .....
\frac{1}{\tilde{x}_{\pi(n)} - \tilde{x}_0}  \; ,
\end{equation}
where the special set of $\tau$ matrices is given by
\[
\{\tau^{(l_0)},\tau^{(l_1)},.....\tau^{(l_n)} \} :=
\]
\begin{equation} \left\{ \;\;
\begin{array}{ll}
\{\tau^{(2)},\tau^{(3)},\tau^{(3)},...\tau^{(3)},\tau^{(1)} \} \;\;\;
\mbox{for}\ n+1 \; = 2m+1 \; \; (n+1\ \mbox{odd)} \\
\{\tau^{(1)},\tau^{(3)},\tau^{(3)},...\tau^{(3)},\tau^{(1)} \} \;\;\;
\mbox{for}\ n+1 \; = 2m+2 \; \; (n+1\ \mbox{even)} \end{array} \right. \;\; .
\end{equation}\\
{\bf Theorem 1 : } \\
For arbitrary $m\geq 1$ :
\begin{equation}
F_{2m+1} \not{\!\!\equiv} 0 \; , \; F_{2m+2} \not{\!\!\equiv} 0 \; .
\end{equation}
{\bf Proof:}
We prove the statement by induction. \\
{\bf 1)} $F_3 \not{\!\!\equiv} 0$ and $F_4 \not{\!\!\equiv} 0$ can
be checked easily.\\
{\bf 2)} Assume $F_{2m+1} \not{\!\!\equiv} 0$ and $F_{2m+2}
\not{\!\!\equiv} 0$ .\\
{\bf 3)} We have to show $F_{2m+3} \not{\!\!\equiv} 0$ and
$F_{2m+4} \not{\!\!\equiv} 0$. First we check $F_{2m+3}$. Again we
abbreviate $2m+3 := n+1$. The trick is to consider
$F_{n+1}(\tilde{x}_0,\tilde{x}_1, ... \tilde{x}_n)$ as a function of
$\tilde{x}_0$, and to compute the corresponding residues
\footnote{We acknowledge a useful discussion with
Peter Breitenlohner on this point.}.
\[
\mbox{Res}_{\tilde{x}_0} [ F_{n+1},\tilde{x}_0 = \tilde{x}_1 ]  = \]
\[ \sum_{\pi(1,2,...n),\pi(1)=1}\!\!\!
\mbox{Tr}\Big[\tau^{(2)} \tau^{(3)} \tau^{(\pi(2))} ...
\tau^{(\pi(n))} \Big]
\frac{1}{\tilde{x}_1\!-\!\tilde{x}_{\pi(2)}}
\frac{1}{\tilde{x}_{\pi(2)}\!-\!\tilde{x}_{\pi(3)}} ...
\frac{1}{\tilde{x}_{\pi(n)}\!-\!\tilde{x}_1}
\]
\begin{equation}
-\!\!\!\!\!\sum_{\pi^\prime(1,2,...n),\pi^\prime(n)=1}\!\!\!
\mbox{Tr}\Big[\tau^{(2)} \tau^{(\pi^\prime(1))}
... \tau^{(\pi^\prime(n-1))} \tau^{(3)} \Big]
\frac{1}{\tilde{x}_1\!-\!\tilde{x}_{\pi^\prime(1)}}
\frac{1}{\tilde{x}_{\pi^\prime(1)}\!-\!\tilde{x}_{\pi^\prime(2)}} ...
\frac{1}{\tilde{x}_{\pi^\prime(n-1)}\!-\!\tilde{x}_1}
\end{equation}
For a given permutation $\pi$ choose the permutation $\pi^\prime$
such that $\pi^\prime(1) = \pi(2),\pi^\prime(2) = \pi(3), ...
\pi^\prime(n-1) = \pi(n), \pi^\prime(n) = \pi(1) = 1$. This gives for
the residue
\[
\sum_{\pi(1,2,...n),\pi(1)=1}
\frac{1}{\tilde{x}_1 - \tilde{x}_{\pi(2)}}
\frac{1}{\tilde{x}_{\pi(2)} - \tilde{x}_{\pi(3)}} ...
\frac{1}{\tilde{x}_{\pi(n)} - \tilde{x}_1}\]
\begin{equation}
\left\{
\mbox{Tr}\Big[\tau^{(2)} \tau^{(3)} \tau^{(\pi(2))} \tau^{(\pi(3))} ...
\tau^{(\pi(n))} \Big] -
\mbox{Tr}\Big[\tau^{(2)} \tau^{(\pi(2))} \tau^{(\pi(3))} ...
\tau^{(\pi(n))} \tau^{(3)} \Big] \right\}  \; .
\end{equation}
Using the cycliticity of the trace and $\{\tau^{(2)},\tau^{(3)}\} = 0$, one
finds that the second trace is the negative of the first one. Furthermore
$\tau^{(2)} \tau^{(3)} = \frac{i}{\sqrt{2}} \tau^{(1)}$.
Looking at the definition
$F_n$ for even $n$ we conclude
\begin{equation}
\mbox{Res}_{\tilde{x}_0} [ F_{n+1},\tilde{x}_0 = \tilde{x}_1 ]  =
i \sqrt{2} F_n(\tilde{x}_1,... \tilde{x}_n) =
i \sqrt{2} F_{2m+2}(\tilde{x}_1,... \tilde{x}_{2m+2}) \; .
\end{equation}
Since by assumption $F_{2m+2} \not{\!\!\equiv} 0$,  Res$[F_{2m+3}]
\neq 0$ and hence $F_{2m+3}$ does not vanish. The same trick can be applied to
prove that this implies $F_{2m+4} \not{\!\!\equiv} 0. \; \Box$

A $n+1$-point function with the flavor content given by (59) is
simply the real part of a multiple of $F_{n+1}$ (compare(56)).
Hence there exist nonvanishing, fully connected $n$-point functions
for arbitrary $n$.

This result implies that it is not possible to bosonize
the whole set of all $N^2$ vector currents $J^{(l)}_\mu$
in terms of free bosons (this was the reason why Witten \cite{witten2}
introduced
his nonabelian bosonization).
This can be done only for a Cartan subalgebra where all generators commute,
and the two traces in equation (62) cancel. The last observation allows to
prove a second theorem. \\
{\bf Theorem 2 :} \\
Any fully connected $n$-point function
\begin{equation}
\Big\langle J^{(1)}_{\mu_0}(x_0) J^{(l_1)}_{\mu_1}(x_1) J^{(l_2)}_{\mu_2}(x_2)
..... J^{(l_n)}_{\mu_n}(x_n) \Big\rangle_c
\end{equation}
containing the $U(1)$ current
vanishes, with the exception of the two point function
$\langle J^{(1)}_\mu(x) J^{(1)}_\nu(y) \rangle_c$
(without loss of generality we shifted
the $U(1)$ current in equation (64) to the first position). \\
{\bf Proof :}
To prove the statement we use the same trick as for theorem 1.
Again we consider the residues of functions $F_{n+1}$ defined by
\[
F_{n+1}(\tilde{x}_0,\tilde{x}_1,... \tilde{x}_n) :=
\]
\begin{equation}
\sum_{\pi(1,2,...n)} \mbox{Tr}\Big[H^{(1)} H^{(l_{\pi(1)})} .....
H^{(l_{\pi(n)})} \Big]
\frac{1}{\tilde{x}_0 - \tilde{x}_{\pi(1)}}
\frac{1}{\tilde{x}_{\pi(1)} - \tilde{x}_{\pi(2)}} .....
\frac{1}{\tilde{x}_{\pi(n)} - \tilde{x}_0}  \; .
\end{equation}
For the residue at $\tilde{x}_0 = \tilde{x}_1$ we obtain an equation
equivalent to (61). After making the same choice for $\pi^\prime$ one
finds
\[
\mbox{Res}_{\tilde{x}_0} [ F_{n+1},\tilde{x}_0 = \tilde{x}_1 ] =
\sum_{\pi(1,2,...n),\pi(1)=1}
\frac{1}{\tilde{x}_1 - \tilde{x}_{\pi(2)}}
\frac{1}{\tilde{x}_{\pi(2)} - \tilde{x}_{\pi(3)}} ...
\frac{1}{\tilde{x}_{\pi(n)} - \tilde{x}_1} \]
\begin{equation}
\left\{
\mbox{Tr}\Big[H^{(1)} H^{(l_1)} H^{(l_{\pi(2)})}  ...
H^{(l_{\pi(n)})} \Big] -
\mbox{Tr}\Big[H^{(1)} H^{(l_{\pi(2)})}  ...
H^{(l_{\pi(n)})} H^{(l_1)} \Big] \right\}  \; .
\end{equation}
Since $H^{(1)}$ commutes with all generators $H^{(l)}$, the two traces
are the same and cancel. Using the same argument one can show that
all residues at $\tilde{x}_1,\tilde{x}_2,.....\tilde{x}_n$ vanish.
So $F_{n+1}$ is analytic and bounded in the entire $\tilde{x_0}$ plane. By
Liouville's theorem $F_{n+1}$ is a constant, and the
limit $\tilde{x}_0 \longrightarrow \infty$ shows that this constant
is zero. Since the $n+1$-point function defined in (64) is
proportional to $F_{n+1}$, it has to vanish. $\Box$

Theorem 2 allows to prove the following proposition about the
structure of the Hilbert space.\\
{\bf Proposition :} \\
The Hilbert space $\cal H$ generated by the vector currents from the
vacuum is the tensor product
\begin{equation}
{\cal H} = {\cal H}_{U(1)} \otimes {\cal H}_{mass=0} \; .
\end{equation}
To prove this we need the connection between untruncated and
fully connected $n$-point functions (see e.g. \cite{glimm}).
\begin{equation}
\langle \phi_1 ..... \phi_n \rangle =
\sum_{\pi \in {\cal P}_n} \prod_{p \in \pi}
\langle \phi_{i_1} ..... \phi_{i_{\mid p \mid}} \rangle_c \; ,
\end{equation}
where ${\cal P}_n$ is the set of all partitions of $\{1,2,...n\}$,
$\pi = \{ p_1, p_2, ... p_{\mid \pi \mid} \}$ denotes an element of
${\cal P}_n$, and $\{i_1,i_2,...i_{\mid p \mid} \}$ is an element $p$ of
$\pi$.
An arbitrary $n+k$-point function (without loss of
generality we write the $U(1)$ currents first;
$l_i \neq 1 , i = 1,2,...k$) factorizes
due to Theorem 2
\[
\Big\langle J^{(1)}_{\mu_1}(x_1) J^{(1)}_{\mu_2}(x_2) .....
J^{(1)}_{\mu_n}(x_n) J^{(l_1)}_{\nu_1}(y_1)
J^{(l_2)}_{\nu_2}(y_2) ..... J^{(l_k)}_{\nu_k}(y_k) \Big\rangle = \]
\[
\sum_{\pi \in {\cal P}_{n+k}} \prod_{p \in \pi}
\Big\langle
J^{(1)}_{\mu_{i_1}}(x_{i_1})
J^{(1)}_{\mu_{i_2}}(x_{i_2}) .....
J^{(1)}_{\mu_{i_{\mid p \mid}}}(x_{i_{\mid p \mid}})
J^{(l_{j_1})}_{\nu_{j_1}}(y_{j_1})
J^{(l_{j_2})}_{\nu_{j_2}}(y_{j_2}) .....
J^{(l_{j_k})}_{\nu_{j_{\mid p \mid}}}(y_{j_{\mid p \mid}}) \Big\rangle_c \; =\]
\[
\left[ \sum_{\pi \in {\cal P}_n} \prod_{p \in \pi}
\Big\langle
J^{(1)}_{\mu_{i_1}}(x_{i_1})
J^{(1)}_{\mu_{i_2}}(x_{i_2}) .....
J^{(1)}_{\mu_{i_{\mid p \mid}}}(x_{i_{\mid p \mid}}) \Big\rangle_c \right]\]
\[
\left[ \sum_{\pi^\prime \in {\cal P}_{k}} \prod_{p^\prime \in \pi^\prime}
\Big\langle
J^{(l_{j_1})}_{\nu_{j_1}}(y_{j_1})
J^{(l_{j_2})}_{\nu_{j_2}}(y_{j_2}) .....
J^{(l_{j_k})}_{\nu_{j_{\mid p\prime \mid}}}
(y_{j_{\mid p^\prime \mid}}) \Big\rangle_c \right] = \]
\begin{equation}
\Big\langle J^{(1)}_{\mu_1}(x_1) J^{(1)}_{\mu_2}(x_2) .....
J^{(1)}_{\mu_n}(x_n) \Big\rangle \;\;
\Big\langle J^{(l_1)}_{\nu_1}(y_1)
J^{(l_2)}_{\nu_2}(y_2) ..... J^{(l_k)}_{\nu_k}(y_k) \Big\rangle \; .
\end{equation}
{}From this the tensor product structure of the Hilbert space follows
easily. $\Box$

We observed in equations (54)-(56) that the dependence of the fully connected
correlations on the gauge field cancels entirely.
Hence the vector currents in the 'massless' sector of the Hilbert space
obey the same algebra as in a system of uncoupled fermions; this is the
well-known level 1 representation of the $SU(N)_L\times SU(N)_R$ current
(Kac-Moody) algebra (see for instance \cite{goddard}).
\section{Confinement}
In order to study confinement, we evaluate the Fredenhagen Marcu
order parameter \cite{fredenhagen}. It is defined by
\begin{equation}
\rho := \lim_{L \rightarrow \infty} \rho(L) :=
\lim_{L \rightarrow \infty} \sum_{\alpha,\beta = 1}^{2}
\frac{{N^{(L)}_{\alpha, \beta}}^\dagger \; N^{(L)}_{\beta, \alpha}}
{\langle W^{(L)} \rangle} \; ,
\end{equation}
where
\begin{equation}
N^{(L)}_{\alpha, \beta} :=
\Big\langle \psi_\alpha^{(a)}(-L,0) U({\cal C}^{(L)})
\overline{\psi}^{(a)}_\beta(L,0) \Big\rangle \;
\mbox{for arbitrary flavor} \; \; a = 1,2,...N \;,
\end{equation}
and
\begin{equation}
U({\cal C}^{(L)}) = e^{ie \int_{{\cal C}^{(L)}} A_\mu(x) dx_\mu} \; .
\end{equation}
For the contour ${\cal C}^{(L)}$ see Fig. 1a.
$W^{(L)}$ denotes the Wilson loop
\begin{equation}
W^{(L)} = e^{ie \int_{{\cal C}^{(L)}_W} A_\mu(x) dx_\mu} \; ,
\end{equation}
along the contour ${\cal C}^{(L)}_W$ given in Fig. 1b. \\
\begin{center}
\unitlength0.6cm
\begin{picture}(19.5,11.5)
\put(1.5,8){\vector(0,1){2}}
\put(1.5,8){\vector(0,-1){1}}
\put(12.5,8){\vector(0,1){2}}
\put(12.5,8){\vector(0,-1){4}}
\put(0.95,8.3){\mbox{L}}
\put(11.6,6.8){\mbox{2L}}
\thicklines
\put(2,7){\line(0,1){3}}
\put(2,10){\line(1,0){6}}
\put(8,7){\line(0,1){3}}
\put(5.2,10.0){\line(-1,1){0.6}}
\put(5.2,10.0){\line(-1,-1){0.6}}
\put(2.3,6.8){\mbox{(-L,0)}}
\put(6.25,6.8){\mbox{(L,0)}}
\put(2,7){\circle*{0.2}}
\put(8,7){\circle*{0.2}}
\put(13,4){\line(0,1){6}}
\put(13,10){\line(1,0){6}}
\put(13,4){\line(1,0){6}}
\put(19,4){\line(0,1){6}}
\put(16.2,10.0){\line(-1,1){0.6}}
\put(16.2,10.0){\line(-1,-1){0.6}}
\put(15.8,4.0){\line(1,1){0.6}}
\put(15.8,4.0){\line(1,-1){0.6}}
\put(13.3,6.8){\mbox{(-L,0)}}
\put(17.25,6.8){\mbox{(L,0)}}
\put(13,7){\circle*{0.2}}
\put(19,7){\circle*{0.2}}
\put(1,2.3){\mbox{{\bf Fig. 1a} Contour for the }}
\put(1,1.5){\mbox{parallel transporter. }}
\put(12,2.3){\mbox{{\bf Fig. 1b} Contour for the }}
\put(12,1.5){\mbox{Wilson loop. }}
\end{picture} \\
\end{center}
The physical interpretation is simple: if there is no confinement
and free `quarks' can be isolated, the free quark lives in a sector orthogonal
to the vacuum sector (obtained by applying gauge invariant operators to the
vacuum); all `mesonic' states belong to the vacuum sector. In particular
$\rho = 0$ follows in this case. In the case of confinement, however,
one expects `fragmentation'. If one tries to separate a
quark-antiquark pair, one is producing another quark-antiquark pair from the
vacuum which binds to the separated pair and forms two mesons, i.e. a state
in the vacuum sector. In this case one expects therefore $\rho\neq 0$.

The computational trick is to rewrite the contour integral over $A_\mu$
as a scalar product with a current having its support on the contour.
e.g.
\begin{equation}
\int_{{\cal C}^{(L)}_W} A_\mu(x) dx_\mu :=
\int A_\mu(x) {j_\mu^W}^{(L)}(x) d^2x \; ,
\end{equation}
with
\begin{equation}
{j^W}^{(L)}(x) = \left( \begin{array}{l}
\theta(x_1 + L) \theta(L-x_1) \; [
\delta(x_2 - L) - \delta(x_2 + L) ] \\
\theta(x_2 + L) \theta(L-x_2) \; [
\delta(x_1 + L) - \delta(x_1 - L) ]
\end{array} \right) \; .
\end{equation}
(strictly speaking a limiting procedure is required here, which we leave as
an exercise).
Performing the functional integration one ends up with
\begin{equation}
\langle W^{(L)} \rangle =
\int d\mu_C[A] e^{ie( A_\mu,{j_\mu^W}^{(L)} )} =
e^{-\frac{e^2}{2}({j_\mu^W}^{(L)}, C_{\mu \nu} {j_\nu^W}^{(L)})} \; .
\end{equation}
For $N_{\alpha \beta}$ in addition a propagator appears in the
gauge field integral
\begin{equation}
N^{(L)}_{\mu \nu} =
\int d\mu_C[A] e^{ie( A_\mu,{j_\mu}^{(L)} )}
G_{\mu \nu}\Big((-L,0),(L,0);A\Big) \; .
\end{equation}
It is useful to rewrite the gauge field integration in terms of the
field $\varphi$ introduced in (16)-(18).
In the chosen representation of the $\gamma$
matrices we have $N^{(L)}_{\mu \mu} = 0$ and $N^{(L)}_{2 1} =
\overline{N}^{(L)}_{1 2}$. Inspecting the quadratic forms that remains after
the functional integration over $\varphi$ one finds
$N^{(L)}_{2 1} = N^{(L)}_{1 2} \in {{I\!\!R}}$. Putting things together
gives
\begin{equation}
\rho^{(L)} = \frac{2}{(2\pi)^2} \frac{1}{(2L)^2} e^{I^{(L)}} \; ,
\end{equation}
where the integral $I^{(L)} := I^{(L)}_s + I^{(L)}_c$ contains two parts
$I^{(L)}_s$ and $I^{(L)}_c$. $I^{(L)}_s$ will be solved explicitely, and
$I^{(L)}_c$ can be shown to go to a nonvanishing
constant for $L \rightarrow \infty$.
$I^{(L)}_s$ is given by $(m^2 = e^2 N/\pi)$
\begin{equation}
I^{(L)}_s = \frac{2e^2}{\pi^2}\int d^2p \frac{1}{(p^2 + m^2)p^2}
\sin^2(p_1 L) \; .
\end{equation}
To solve it, we transform it to polar coordinates and integrate over
the angle. This leads to an integral over a Bessel function, which
can (after some partial integration) be found in integral tables.
The result is
\begin{equation}
I^{(L)}_s = 2\frac{1}{N}\ln(2L) +
\frac{1}{N}K_0\left(2Le\sqrt{\frac{N}{\pi}}\right) +
\frac{1}{N}\ln\left(\frac{e}{2}\sqrt{\frac{N}{\pi}}\right) +
\frac{\gamma}{N} \; ,
\end{equation}
where $\gamma$ denotes Euler's constant. The modified Bessel function $K_0$
vanishes for $L \rightarrow \infty$.
The second integral $I^{(L)}_c$ is given by
\[
I^{(L)}_c = \frac{e^2}{\pi^2}\int d^2p \frac{1}{p^2 + m^2} \Big\{ -
\frac{\sin^2(p_1 L)}{(p_1)^2} + \]
\[
4\sin^2(p_1 L) \frac{1}{(p_2)^2} [
\sin^2(p_2 L) - \sin^2(p_2 L/2) ] \Big\} \; = \]
\[ \frac{e^2}{\pi^2}\int d^2p \frac{1}{(p^2 + m^2)(p_2)^2} \Big\{
\sin^2(p_2 L)  - 2\sin^2(p_2 L/2) \Big\} \; - \]
\begin{equation}
\frac{2e^2}{\pi^2}\int d^2p \frac{1}{(p^2 + m^2)(p_2)^2} \cos(p_1 2L) \Big\{
\sin^2(p_2 L)  - \sin^2(p_2 L) \Big\} \; .
\end{equation}
Performing the $p_1$ integration in the second term, one sees that
this integral is exponentially decaying for $L \rightarrow \infty$.
The first term we integrate over $p_2$ first and find the cancellation
of two terms linear in $L$. The remaining part gives an exponentially
suppressed term, and the constant $1/N$.

Putting things together, we obtain for $\rho(L)$
\begin{equation}
\rho(L) = \frac{1}{2\pi^2}
\frac{e^2}{4}\left(\frac{N}{\pi} \right)^{\frac{1}{N}}
e^{\frac{1}{N}(2 \gamma + 1)} \left(\frac{1}{2L}\right)^{2\frac{N-1}{N}} \; .
\end{equation}
(In the exponent we dropped terms that vanish for $L \rightarrow \infty$.)
For the confinement parameter this gives two different results depending
on the number of flavors
\begin{equation}
\rho = \lim_{L \rightarrow \infty} \rho(L) =
\left\{ \begin{array}{cr}
\frac{1}{2\pi^2} \frac{e^2}{4}\frac{1}{\pi}
e^{2\gamma +1} & \mbox{for}\;\; N = 1 \\
0 & \mbox{for}\;\; N > 1
\end{array} \right. \; . \end{equation}
This result may appear surprising. Confinement of a quark-antiquark
system depends on the number of flavors! (Up to a factor the $N=1$ result
can be found in \cite{alonso}.)

Having solved the above case one can easily generalize the formula
to products of $n$ spinors with pairwise distinct flavor quantum
numbers. Define for $n \leq N$
\begin{equation}
N^{(L)}_{\alpha_1 ... \alpha_n \beta_1 ... \beta_n} :=
\Big\langle \psi^{(1)}_{\alpha_1}(-L,0) ...
\psi^{(n)}_{\alpha_n}(-L,0) U({\cal C}^{(L)})^n
\overline{\psi}^{(1)}_{\beta_1}(L,0) ...
\overline{\psi}^{(n)}_{\beta_n}(L,0) \Big\rangle \; .
\end{equation}
The corresponding $\rho^{(n)}(L)$ is given by
\begin{equation}
\rho^{(n)}(L) := \frac{\sum_{\alpha_i,\beta_j}
{N^{(L)}}^\dagger_{\alpha_1 ... \alpha_n \beta_1 ... \beta_n}
N^{(L)}_{\beta_1 ... \beta_n \alpha_1 ... \alpha_n}}
{\langle {W^{(L)}}^n \rangle} \; =
\end{equation}
\begin{equation}
\frac{2}{(4\pi L)^{2n}} e^{n^2 I^{(L)}} + O(1/L) =
\frac{2}{(2\pi)^{2n}} \left(\frac{e^2 N}{4\pi}\right)^{\frac{n^2}{N}}
e^{\frac{n^2}{N}(2\gamma +1)}
{\left(\frac{1}{2L}\right)}^{2\frac{nN - n^2}{N}} \; + O(1/L) \; .
\end{equation}
Performing the limit $L \rightarrow \infty$ we obtain
\begin{equation}
\rho^{(n)} = \lim_{L \rightarrow \infty} \rho^{(n)}(L) =
\left\{ \begin{array}{cr}
\frac{2}{(2\pi)^{2N}} \Big(\frac{e^2 N}{4 \Pi}\Big)^{N}
e^{N(2\gamma +1)} & \mbox{for}\;\; n = N \\
0 & \mbox{for}\;\; n < N
\end{array} \right. \; . \end{equation}
Thus in the $N$-flavor model an arrangement of $N$ `quarks' is bound by
a confining force to an arrangement of $N$ `antiquarks'.

A further generalization shows that an operator of $N + n, n \leq N$
quarks behaves like the product of only $n$ quarks.
The corresponding $N$ is defined by
\[
N_{\alpha_1...\alpha_N \alpha^\prime_1...\alpha^\prime_n
\beta_1...\beta_N \beta^\prime_1...\beta^\prime_n}^{(L)} := \]
\begin{equation}
\Big\langle\psi^{(1)}_{\alpha_1} ...
\psi^{(N)}_{\alpha_N}
\psi^{(1)}_{\alpha^\prime_1} ...
\psi^{(n)}_{\alpha^\prime_n}(-L,0) U({\cal C}^{(L)})^{N+n}
\overline{\psi}^{(1)}_{\beta_1} ...
\overline{\psi}^{(N)}_{\beta_N}
\overline{\psi}^{(1)}_{\beta^\prime_1} ...
\overline{\psi}^{(n)}_{\beta^\prime_n}(L,0) \Big\rangle \; .
\end{equation}
For the fields with flavors $1,2,...n$ extra contractions are possible.
Some of these cancel each other. The remaining terms have no dependence
on $\varphi$ and contribute only to the
$1/L$ behaviour of the free propagator. The
flavors $n+1,n+2,...N$ are seen to contribute a factor
\begin{equation}
e^{(N-n)^2 \; I^{(L)} } \;,
\end{equation}
after performing the path integral. Collecting all terms one obtains
\begin{equation}
\rho^{(N+n)} = \lim_{L \rightarrow \infty} \rho^{(N+n)}(L) =
\left\{ \begin{array}{cr}
\mbox{const} \; \; \neq 0 & \mbox{for}\;\; n = 0,N \\
0 & \mbox{for}\;\; 0 < n < N
\end{array} \right. \; . \end{equation}
The physical interpretation suggested by this behavior of the
Fredenhagen-Marcu order parameters is the following: The model has $N$
distinct superselection sectors labeled by a charge $Q$ that is defined
only modulo $N$. To obtain a state in the the sector of charge $Q=n$,
$n<N$, one applies an operator consisting of $n$ `quarks' and $n$
antiquarks, separated by distance $L$ and takes the limit $L\to\infty$.

\section{Decomposition into clustering states and the vacuum angle}
The state (expectation functional) which we have constructed so far
violates clustering, as we will show in this section. We will also
construct an integral decomposition of this state into `pure phases'
labeled by a vacuum angle $\theta$ and satisfying the cluster decomposition
property. Thus we recover the structure that is usually associated
with the existence of topological sectors of the gauge field. The words
`pure phase' should not be taken literally, because the different states
are not related by a symmetry operation as in the case of spontaneous
symmetry breaking. The transformations that intertwine the different
$\theta$-states -- the axial $U(1)$ transformations -- are not symmetry
transformations because the corresponding current is not conserved (due to
the presence of the Adler-Bardeen anomaly), and therefore the corresponding
Ward identities contain an anomaly term.

\subsection{Identification of operators that violate clustering}
To identify the operators that violate clustering, we start with an
ansatz containing only the chiral densities
$\overline{\psi}^{(a)} P_\pm \psi^{(a)} \;\; ( P_\pm := (1 \pm \gamma_5)/2 )$
and discuss the effect of
adding vector currents later. Define
\[
C(\tau) = \Big\langle \prod_{a=1}^N
\prod_{i=1}^{n_a} \overline{\psi}^{(a)}(x^{(a)}_i\!+ \hat{\tau})
P_+ \psi^{(a)}(x^{(a)}_i\!+ \hat{\tau})
\prod_{i=1}^{m_a} \overline{\psi}^{(a)}(y^{(a)}_i\!+ \hat{\tau})
P_- \psi^{(a)}(y^{(a)}_i\! + \hat{\tau}) \]
\[
\prod_{i=1}^{n^\prime_a}
\overline{\psi}^{(a)}({x^\prime}^{(a)}_i)
P_+ \psi^{(a)}({x^\prime}^{(a)}_i)
\prod_{i=1}^{m^\prime_a}
\overline{\psi}^{(a)}({y^\prime}^{(a)}_i)
P_- \psi^{(a)}({y^\prime}^{(a)}_i) \Big\rangle \; - \]
\[
\Big\langle \prod_{a=1}^N
\prod_{i=1}^{n_a} \overline{\psi}^{(a)}(x^{(a)}_i )
P_+ \psi^{(a)}(x^{(a)}_i )
\prod_{i=1}^{m_a} \overline{\psi}^{(a)}(y^{(a)}_i )
P_- \psi^{(a)}(y^{(a)}_i ) \Big\rangle \]
\begin{equation}
\Big\langle \prod_{a=1}^N
\prod_{i=1}^{n^\prime_a}
\overline{\psi}^{(a)}({x^\prime}^{(a)}_i)
P_+ \psi^{(a)}({x^\prime}^{(a)}_i)
\prod_{i=1}^{m^\prime_a}
\overline{\psi}^{(a)}({y^\prime}^{(a)}_i)
P_- \psi^{(a)}({y^\prime}^{(a)}_i) \Big\rangle \; ,
\end{equation}
where $\hat{\tau}$ is the vector of length $\tau$ in 2-direction.
Violation of the cluster property now manifests itself in a nonvanishing
limit
\begin{equation}
\lim_{\tau \rightarrow \infty} C(\tau) =: C  \neq 0 \; .
\end{equation}
Because $G_{\alpha,\alpha} = 0$, the second term contributes
only if $m_a = n_a , m^\prime_a = n^\prime_a$ for all $a=1,...N$,
and the first one only if
\begin{equation}
n_a + n^\prime_a = m_a + m^\prime_a \; \; , \; \; a = 1,...N \;.
\end{equation}
After some reordering the first term reads
\[ \Big \langle
\prod_{a=1}^N
\prod_{i=1}^{n_a} \psi^{(a)}_1(x_i^{(a)}\!+ \hat{\tau})
\prod_{i=1}^{m^\prime_a} \overline{\psi}^{(a)}_2({y^\prime}^{(a)}_i)
\prod_{i=1}^{n^\prime_a} \psi^{(a)}_1({x^\prime}_i^{(a)})
\prod_{i=1}^{m_a} \overline{\psi}^{(a)}_2(y_i^{(a)}\!+ \hat{\tau}) \]
\begin{equation}
\prod_{i=1}^{m^\prime_a} \psi^{(a)}_2({y^\prime}_i^{(a)})
\prod_{i=1}^{n_a} \overline{\psi}^{(a)}_1(x_i^{(a)}\!+ \hat{\tau})
\prod_{i=1}^{m_a} \psi^{(a)}_2(y_i^{(a)}\!+ \hat{\tau})
\prod_{i=1}^{n^\prime_a} \overline{\psi}^{(a)}_1({x^\prime}^{(a)}_i)
\Big \rangle \; .
\end{equation}
We dropped an overall sign depending on $n_a,m_a,n^\prime_a,m^\prime_a$
which is not relevant yet. Using the explicit form (20) of the propagator,
one finds that the gauge field integral can be factorized off:
\begin{equation}
\langle \cdot \rangle = I \; \langle \cdot \rangle_0 \;,
\end{equation}
where $\langle \cdot \rangle$ stands for the expectation value
(94), and $\langle \cdot \rangle_0$ denotes the same
expectation value for $e = 0$. The factor $I$ is given by
\begin{equation}
I = \int d\mu_{\tilde{C}}[\varphi] e^{2e\sum_{a=1}^N \Big[
\sum_{i=1}^{m_a} \varphi(y^{(a)}_i\!+ \hat{\tau}) +
\sum_{i=1}^{m^\prime_a} \varphi({y^\prime}^{(a)}_i) -
\sum_{i=1}^{n_a} \varphi(x^{(a)}_i\!+ \hat{\tau}) -
\sum_{i=1}^{n^\prime_a} \varphi({x^\prime}^{(a)}_i) \Big]} \; .
\end{equation}
It has the general structure
\[
\int d \mu_{\tilde{C}}[\varphi]
e^{2 e\sum_{i=1}^{m} [
\varphi(w_i) - \varphi(z_i) ] } \; = \]
\begin{equation}
e^{\sum_{i,j}^m V(w_i -  z_j)
- \frac{1}{2}\sum_{i \neq j}^m V(w_i - w_j)
- \frac{1}{2}\sum_{i \neq j}^m V(z_i - z_j) } \; ,
\end{equation}
where we performed the functional integral and used (19)
for the covariance $\tilde{C}$, to obtain after some reordering of the
arguments
\begin{equation}
V(x) = 4 e^2 \int \frac{d^2p}{(2\pi)^2}\frac{1-\cos(px)}{p^2(p^2+m^2)} \; .
\end{equation}
Up to a trivial factor this is the integral defined in equation (79),
which we already
solved in the last section. The result is
\begin{equation}
V(x) = \frac{1}{N}\ln (x^2) + V^\prime(x) , \; \;
V^\prime(x) = \frac{2}{N} \left[
\mbox{K}_0\left(e\sqrt{\frac{N}{\pi}}|x|\right) +
\ln\left(\frac{e}{2}\sqrt{\frac{N}{\pi}}\right) + \gamma \right] \; .
\end{equation}
where we split $V(x)$ into the logarithmic part and a
part $V^\prime(x)$ which approaches the constant
$\frac{2}{N}[\ln(\frac{e}{2}\sqrt{\frac{N}{\pi}}) + \gamma ]$
for $x \rightarrow \infty$.

When applying (97) to (96), the sets $\{w_i\}$ and $\{z_i\}$ are
given by
\[
w_i^{(a)} \in
\{ y_l^{(b)} +\hat{\tau} , {y^\prime}^{(a)}_k \; | \;
l = 1,...m_a ; k = 1,...m^\prime_a ; b = 1,...N \} \; ,
\]
\begin{equation}
z_i^{(a)} \in
\{ x_l^{(b)} +\hat{\tau} , {x^\prime}^{(a)}_k \; | \;
l = 1,...n_a ; k = 1,...n^\prime_a ; b = 1,...N \} \; .
\end{equation}
Inserting these arguments in (97) and collecting the terms with
nontrivial $\tau$ dependence, we obtain
for the large-$\tau$ behaviour of the gauge field integral a factor
\begin{equation}
\Big( \tau^2 \Big)^{ \frac{1}{N} \Big[
\big(\sum_a n_a\big) \big(\sum_a m^\prime_a\big) +
\big(\sum_a m_a\big) \big(\sum_a m^\prime_a\big) -
\big(\sum_a n_a\big) \big(\sum_a n^\prime_a\big) -
\big(\sum_a n_a\big) \big(\sum_a m^\prime_a\big) \Big]} \; .
\end{equation}
The $\langle \cdot \rangle_0$ factor is
\[
\Big \langle \cdot \Big \rangle_0 =
\prod_{a=1}^N
\Big \langle
\prod_{i=1}^{n_a} \psi^{(a)}_1(x_i^{(a)}\!+ \hat{\tau})
\prod_{i=1}^{m^\prime_a} \overline{\psi}^{(a)}_2({y^\prime}^{(a)}_i)
\prod_{i=1}^{n^\prime_a} \psi^{(a)}_1({x^\prime}_i^{(a)})
\prod_{i=1}^{m_a} \overline{\psi}^{(a)}_2(y_i^{(a)}\!+ \hat{\tau})
\]
\begin{equation}
\prod_{i=1}^{m^\prime_a} \psi^{(a)}_2({y^\prime}_i^{(a)})
\prod_{i=1}^{n_a} \overline{\psi}^{(a)}_1(x_i^{(a)}\!+ \hat{\tau})
\prod_{i=1}^{m_a} \psi^{(a)}_2(y_i^{(a)}\!+ \hat{\tau})
\prod_{i=1}^{n^\prime_a} \overline{\psi}^{(a)}_1({x^\prime}^{(a)}_i)
\Big \rangle_0 \; .
\end{equation}
It can be expressed in terms of determinants. The general structure is
\[ C_0 :=
\Big \langle
\prod_{i=1}^n \psi_1(z_i)
\overline{\psi}_2(w_i)
\prod_{i=1}^n \psi_1(z_i)
\overline{\psi}_2(w_i) \Big\rangle_0 \; =
\]
\[
\left( \frac{1}{2\pi} \right)^{2n}
\sum_{\pi(n)} \mbox{sign}
\frac{1}{\tilde{z}_1 - \tilde{w}_{\pi(1)}} ...
\frac{1}{\tilde{z}_n - \tilde{w}_{\pi(n)}}
\overline{\sum_{\pi^\prime(n)} \mbox{sign}
\frac{1}{\tilde{z}_1 - \tilde{w}_{\pi^\prime(1)}} ...
\frac{1}{\tilde{z}_n - \tilde{w}_{\pi^\prime(n)}}} \; =\]
\begin{equation}
\left( \frac{1}{2\pi} \right)^{2n} \left|
{ \; \atop { \mbox{det} \atop {\scriptstyle (i,j)} } }\Big(
\frac{1}{\tilde{z}_i - \tilde{w}_j} \Big) \right|^2 \; .
\end{equation}
Determinants of this type can be rewritten using Cauchy's identity
(see e.g. \cite{deutsch})
\begin{equation}
{ \; \atop { \mbox{det} \atop {\scriptstyle (i,j)} } }\Big(
\frac{1}{\tilde{z}_i - \tilde{w}_j} \Big) =
(-1)^{\frac{n(n-1)}{2}}
\frac{\prod_{1 \leq i < j \leq n} (\tilde{z}_i - \tilde{z}_j )
( \tilde{w}_i - \tilde{w}_j ) }
{\prod_{i,j = 1}^n ( \tilde{z}_i - \tilde{w}_j ) }\; .
\end{equation}
Hence we obtain
\begin{equation}
C_0 =
\frac{\prod_{1 \leq i < j \leq n} (z_i - z_j )^2 ( w_i - w_j )^2 }
{\prod_{i,j = 1}^n ( z_i - w_j )^2 } \; .
\end{equation}
Inserting the sets (100) for $w_i,z_i$ and collecting the
$\tau$ dependent powers gives the large $\tau$ behaviour
\begin{equation}
\prod_{a=1}^N \Big(\tau^2\Big)^{
n_a n^\prime_a + m_a m^\prime_a -
n_a m^\prime_a - m_a n^\prime_a} =
\Big(\tau^2\Big)^{\sum_a (n_a - m_a)(n^\prime_a - m^\prime_a) } \; .
\end{equation}
Combining this with (101) gives the total large-$\tau$ dependence for
the first term of $C(\tau)$
\begin{equation}
\Big(\frac{1}{\tau^2}\Big)^{ \frac{1}{N} \sum_{a,b=1}^N
(n_a-m_a)(n^\prime_b - m^\prime_b) - \sum_{a=1}^N
(n_a-m_a)(n^\prime_a - m^\prime_a)} =:
\Big(\frac{1}{\tau^2}\Big)^E \; .
\end{equation}
Violation of clustering can only appear if the exponent vanishes and
$C(\tau)$ cannot approach zero.
Inserting the condition (93) we can rewrite the exponent
\[
E =  \sum_{a=1}^N (n_a\!-\!m_a)(n_a\!-\!m_a) -
\frac{1}{N} \sum_{a,b=1}^N
(n_a\!-\!m_a)(n_b\!-\!m_b) = \]
\begin{equation}
\frac{1}{N}  \sum_{a,b=1}^N
(n_a\!-\!m_a)R_{ab}(n_b\!-\!m_b) \; ,
\end{equation}
where the matrix $R$ was already defined in (25), section 3.
There we solved the corresponding eigenvalue problem.
We found one eigenvalue 0, and $N-1$ eigenvalues $N$. The eigenvector
$x^0$ to the eigenvalue 0 is given by $x^0 = 1/\sqrt{N} (1,1,...1)^T$.
Hence the quadratic form $x^T R x$ is positive semidefinite, and
vanishes only if $x$ is a multiple of $x^0$. This implies that the
exponent $E$ is nonnegative and vanishes only for
\begin{equation}
n_a - m_a = m^\prime_a - n^\prime_a = n  \;\; , \;\;  n \in
\mbox{Z\hspace{-.8ex}Z} \, .
\end{equation}
In the special case $n=0$ the second term in (91) does not
vanish. It cancels the contribution of the first term. Hence
we find no violation of clustering for $n=0$. As already mentioned,
for $n\neq 0$ the second term vanishes due to $G_{\alpha \alpha} = 0$,
and no such cancellation is possible.

How does this picture change when we allow vector currents as well?
First we notice that vector currents do not contribute to the
gauge field integral. Consider e.g. the term
$\overline{\psi}^{(1)}_1(x) \psi^{(2)}_2(x)$. The $\overline{\psi}^{(1)}_1(x)$
generates a propagator $G_{21}(\cdot,x)$, the $\psi^{(2)}_2 (x)$ a propagator
$G_{21}(x,\cdot)$. Inspecting (20) immediately shows the cancellation
of the $\varphi$ dependence in the product of the
propagators. Hence each vector current can only
contribute a $1/\tau$ from the free propagator.

Nonvanishing results remain only if the flavors that occur in the
vector currents can contract entirely. Thus we have to consider
only `closed cycles' like e.g.
\begin{equation}
\overline{\psi}^{(1)}(x) \gamma_\mu \psi^{(2)}(x) \;\;\;
\overline{\psi}^{(2)}(y) \gamma_\nu \psi^{(3)}(y) \;\;\;
\overline{\psi}^{(3)}(z) \gamma_\omega \psi^{(1)}(z) \; .
\end{equation}
In principle there are two possibilities to distribute the space-time
arguments $x,y,z$. If they are all in one cluster they do not bring in
any $\tau$ dependence. They do not modify the clustering, only the constant
$C$. If one distributes the closed cycle over both clusters
then the situation changes. Each vector current with a partner in the
other cluster contributes a $1/\tau$ from the free propagator.
This implies that any combination of vector currents alone clusters.
Nevertheless a combination
of vector currents together with the ansatz (91) could violate clustering.
But we saw that the gauge field integral contributions from the chiral
charges
$\overline{\psi} P_\pm \psi$ can at most compensate the $1/\tau$
of these charges, nothing else
($x^T R x$ is positive semidefinite, compare (107)).
Hence adding vector currents that can only contract between the clusters
can at most enforce operators to cluster, never create extra
powers of $\tau$  that lead to violation of clustering. The same is true
when inserting currents of the Cartan subalgebra where one has to introduce
a point splitting regulator. The gauge field transporter that
connects $\overline{\psi}(x-\varepsilon)$ and $\psi(x+\varepsilon)$
only is a modification within a cluster that does not change the
clustering behaviour.

We now want to analyze the symmetry properties of the nonclustering
operators.
To make our notation more convenient we introduce
\begin{equation}
{\cal O}_\pm ( \{x\} ) :=
\prod_{a=1}^{N}\overline{\psi}^{(a)}(x^{(a)})P_\pm \psi^{(a)}(x^{(a)}) \; .
\end{equation}
It will turn out that the lack of clustering of ${\cal O}_\pm$
is related to the fact that they
are singlets under the conserved symmetry group
$U(1)_V \times SU(N)_L\times SU(N)_R$, but
transform nontrivially under the explicitly broken $U(1)_A$.
To obtain operators that transform under a single irreducible representation
of the symmetry group, we antisymmetrize ${\cal O}_\pm$ with respect to the
flavor indices and call the result ${\cal O}_\pm^a$.
\[
{\cal O}_\pm^a( \{x\} ) := (-1)^{\frac{N(N-1)}{2}}\left[
\frac{1}{N!} \sum_\pi \mbox{sign}(\pi) \prod_{a=1}^{N}
\overline{\psi}^{(\pi(a))}_{\stackrel{{\scriptstyle 1}}{{\scriptstyle 2}}}
(x^{(a)})
\right] \]
\begin{equation}
\left[ \frac{1}{N!} \sum_{\pi^\prime} \mbox{sign}(\pi^\prime)
\prod_{a^\prime=1}^{N}
\psi^{(\pi^\prime(a^\prime))}_{\stackrel{{\scriptstyle 1}}{{\scriptstyle 2}}}
(x^{(a^\prime)}) \right] \; .
\end{equation}
The global sign comes from shifting all $\overline{\psi}$ to the left.
Using
\begin{equation}
\prod_{a=1}^N \overline{\psi}^{(a)}_\alpha(x^{(\pi(a))}) =
\mbox{sign}(\pi^{-1}) \prod_{a=1}^N
\overline{\psi}^{(\pi^{-1}(a))}_{\alpha}(x^{(a)})
\; ,
\end{equation}
one can express ${\cal O}^a_\pm$ as well in terms of symmetrized
space time arguments,
\begin{equation}
{\cal O}_\pm^a( \{x\} ) := \frac{1}{(N!)^2}\sum_{\pi,\pi^\prime}
\prod_{a=1}^{N}\overline{\psi}^{(a)}(x^{(\pi(a))})
P_\pm \psi^{(a)}(x^{(\pi(a))}) \; .
\end{equation}
Since the constant $C := \lim_{\tau \rightarrow \infty} C(\tau)$
is invariant under
the permutation of arguments within a cluster (compare (97),(104)), the
latter expression shows explicitely that the constant $C$
is the same for $\cal O_\pm$ and ${\cal O}^a_\pm$.
{}From (112) one easily reads off the invariance of ${\cal O}^a_\pm$ under
\begin{equation}
U(1)_V \times SU(N)_L \times SU(N)_R \; .
\end{equation}
We end up with the following picture of the clustering problem:
The prototype of a correlation function that violates clustering is given by
\[
C(\tau) =
\Big \langle
\prod_{i=1}^n {\cal O}^a_+ ( \{ x_i + \hat{\tau} \} )
\prod_{i=1}^m {\cal O}^a_- ( \{ y_i + \hat{\tau} \} )
\prod_{i=1}^{n^\prime} {\cal O}^a_+ ( \{ x^\prime_i \} )
\prod_{i=1}^{m^\prime} {\cal O}^a_- ( \{ y^\prime_i \} ) \Big\rangle - \]
\begin{equation}
\Big\langle
\prod_{i=1}^n {\cal O}^a_+ ( \{ x_i \} )
\prod_{i=1}^m {\cal O}^a_- ( \{ y_i \} ) \Big\rangle \;\; \Big\langle
\prod_{i=1}^{n^\prime} {\cal O}^a_+ ( \{x^\prime_i \} )
\prod_{i=1}^{m^\prime} {\cal O}^a_- ( \{y^\prime_i \} ) \Big\rangle \; ,
\end{equation}
with the condition
\begin{equation}
n-m = -n^\prime + m^\prime \in \mbox{Z\hspace{-.8ex}Z}\backslash
\{0\} \; .
\end{equation}
Insertion of closed cycles of vector currents into a
cluster does not change the
clustering behaviour. If one inserts vector currents that can contract
only to a partner in the other cluster,
operators that violated clustering before now become operators
obeying the
cluster property. Of course it is possible to generalize
the operators ${\cal O}_\pm$
further by e.g. splitting the arguments and connect them with a
parallel transporter. Since this is a modification within a
cluster, the extra terms in the functional integral will not
depend on $\tau$ and only modify the constant.

For completeness we quote the constant $C = \lim_{\tau \rightarrow \infty}
C(\tau)$ with $C(\tau)$ defined in (116)
\begin{equation}
C =  {\cal F} (\{x_i\},\{y_i\}) \; \;
{\cal F} ( \{x^\prime_i\},\{y^\prime_i\}) \; .
\end{equation}
The dependence on the space time arguments factorizes into two parts
that depend on the arguments in the two clusters. This function
$\cal F$ is unique for an operator, if only the other operator has the
right quantum numbers to form a pair that violates clustering. It is
given by
\[
{\cal F}(\{x_i\},\{y_i\}) = \prod_{a=1}^N
\frac{\prod_{1 \leq i < j \leq n} \left(x_i^{(a)} - x_j^{(a)}\right)^2
\prod_{1 \leq i < j \leq m} \left(y_i^{(a)} - y_j^{(a)}\right)^2}
{\prod_{i=1}^n \prod_{j=1}^m \left(x_i^{(a)} - y_j^{(a)} \right)}
\]
\begin{equation}
e^{\sum V \left(x_i^{(a)} - y_j^{(b)} \right)
-\frac{1}{2}\sum V \left(x_i^{(a)} - x_j^{(b)} \right)
-\frac{1}{2}\sum V \left(y_i^{(a)} - y_j^{(b)} \right) }
\left( \frac{1}{2\pi} \right)^{N(n+m)}
\left[\frac{e^2 N}{4 \pi} e^{2\gamma} \right]^{\frac{N(n-m)^2}{2}} \; \;
\; .
\end{equation}
$\sum$ denotes summation over all possible values of the indices $a,b,i,j$.
This factorization property remains valid when vector currents
are inserted.

\subsection{Decomposition into clustering states}
To decompose the vacuum state of the theory in terms of clustering
$\theta$ vacua we
use the charge that is associated to the axial $U(1)$ transformation
\begin{equation}
\psi^{(a)} \; \longrightarrow \; e^{i \varepsilon \gamma_5}
\psi^{(a)} \; .
\end{equation}
An arbitrary product $\cal B$ of $\overline{\psi}_\alpha^{(a)}(x),
\psi^{(a^\prime)}_{\alpha^\prime}(x^\prime)$ transforms under $U(1)_A$
\begin{equation}
{\cal B}({\scriptstyle\{x\}}) \; \longrightarrow
e^{ i m} {\cal B}({\scriptstyle\{x\}}) \; , \; m \in
\mbox{Z\hspace{-.8ex}Z} \; .
\end{equation}
(More generally we can consider observables that are sums of operators
with definite transformation properties under $U(1)_A$).
Define the corresponding charge $Q_5({\scriptstyle {\cal B}})$ as $m$.
Obviously ${\cal O}_\pm$ have the charge $\pm 2N$.
We will now decompose the expectation functional $\langle \cdot\rangle$
into states $\langle \cdot\rangle_\theta$ labeled by a parameter
$\theta \in [-\pi,\pi]$ defined as follows
\begin{equation}
\langle {\cal B}({\scriptstyle\{x\}}) \rangle_\theta :=
e^{i \theta Q_5({\scriptstyle{\cal B}})} \lim_{\tau \rightarrow \infty}
\langle {\cal U}_\tau ({\scriptstyle {\cal B}}) \;
{\cal B}({\scriptstyle\{x\}}) \rangle \; .
\end{equation}
The set of `test operators' ${\cal U}_\tau ({\scriptstyle {\cal B}})$
is defined by
\begin{equation}
{\cal U}_\tau ({\scriptstyle {\cal B}}) :=
\left\{ \begin{array}{l}
{\cal N}^{(n)} (\{x\})  \prod_{i=1}^n {\cal O}^a_\mp (\{x_i\})
\; \mbox{for} \; \; Q_5({\scriptstyle{\cal B}}) = \pm 2nN , n \geq 1 \; , \\
1 \; \mbox{otherwise} \; \; . \end{array} \right.
\end{equation}
Up to the requirement of being nondegenerate,
the arguments $\{x\}$ are arbitrary.
The normalizing factor ${\cal N}^{(n)} (\{x\})$ is defined such that \\
$\lim_{\tau^\prime \rightarrow \infty}
\langle {\cal U}_{\tau^\prime} ({\scriptstyle{\cal B}^\dagger})
\; {\cal U}_\tau ({\scriptstyle{\cal B}}) \rangle = 1$.
It can be read off from (119)
\begin{equation}
{\cal N}^{(n)} (\{x\}) =
\left( \frac{1}{2\pi} \right)^{-Nn} \!
\left[\frac{e^2 N}{4 \pi} e^{2\gamma} \right]^{-\frac{Nn^2}{2}}
\prod_{a=1}^N \prod_{1 \leq i < j \leq n} \!
\left(x_i^{(a)}\!-\!x_j^{(a)}\right)^{-2}
e^{\frac{1}{2}  V\left(x_i^{(a)} - x_j^{(b)} \right) } \; .
\end{equation}
The expectation functionals (states) $\langle\cdot\rangle_\theta$ have
the following properties:\\
{\bf Theorem 3 :}\\
{\bf i) } The state $\langle \cdot\rangle$ constructed initially
is recovered by averaging over $\theta$
\begin{equation}
\langle \; \cdot \; \rangle = \frac{1}{2\pi}\int_{-\pi}^{\pi}
\langle \; \cdot \; \rangle_\theta d\theta \;.
\end{equation}
{\bf ii) } The cluster decomposition property holds.\\
 \\
{\bf Proof:} \\
{\bf i):}
The averaging procedure leaves a nonvanishing result only for
operators $\cal B$ with vanishing charge $Q_5({\scriptstyle {\cal B}}) = 0$
\begin{equation}
\langle {\cal B} \rangle = \frac{1}{2\pi}\int_{-\pi}^\pi
\langle {\cal B} \rangle_\theta d\theta =
\frac{1}{2\pi}\int_{-\pi}^\pi \; e ^{i \theta Q_5({\scriptstyle{\cal B}}) } \;
d\theta \;  \lim_{\tau \rightarrow \infty}
\langle {\cal U}_\tau({\scriptstyle {\cal B}}) \;
{\cal B} \rangle =
\delta_{Q_5({{\scriptscriptstyle \cal B}}),0} \; \langle {\cal B} \rangle \; .
\end{equation}
In the last step we used ${\cal U}_\tau({\scriptstyle {\cal B}}) = 1$
if $Q_5({\scriptstyle {\cal B}}) = 0$. To complete the proof,
one has to check that in the state $\langle\cdot\rangle$
constructed initially the vacuum expectation values of operators $\cal B$ with
$Q_5({\scriptstyle {\cal B}}) \neq 0$ vanish (cf. also \cite{patr}).
In our representation of the
$\gamma$ matrices this can be seen immediately.
$Q_5({{\scriptstyle\cal B}}) \neq 0$
means that the number of $\overline{\psi}_1,\psi_1$ is not equal to the
number $\overline{\psi}_2,\psi_2$. Since $G_{\alpha \alpha} = 0$, for
each $\psi_1$ there has to be a $\overline{\psi}_2$ and for each
$\psi_2$ a $\overline{\psi}_1$ to give a nonvanishing contribution.
But this implies that the number of fields
$\overline{\psi}_1,\psi_1$ in $\cal B$
is equal to the number of fields $\overline{\psi}_2,\psi_2$.
Hence
\begin{equation}
\langle {\cal B} \rangle = 0  \;\; \mbox{for} \;\;
Q_5({\scriptstyle {\cal B}}) \neq 0 \;.
\end{equation}

\noindent
{\bf ii):}
Let $\cal A$ and $\cal B$ be arbitrary operators. Define
\begin{equation}
C_\theta(\tau) := \lim_{\tau \rightarrow \infty} \left[
\langle {\cal A}(\tau) {\cal B}(0) \rangle_\theta \; - \;
\langle {\cal A}(0) \rangle_\theta \;
\langle {\cal B}(0) \rangle_\theta \right] \; .
\end{equation}
We do not display the dependence on the
space time arguments $\{ x \}$ explicitely,
only the dependence on the shift variable $\tau$.
Depending on the axial charge $Q_5$ of the operators
$\cal A,B,AB$, we have to insert the different definitions of
$\langle \cdot \rangle_\theta$.
We introduce the following convenient notation:
An operator for which the first alternative in equation (123)
holds we call a `type {\bf I}' operator, the operators where the second
alternative holds is called `type {\bf II}'. We remark that all operators of
type {\bf II } cannot have the problem of violating the clustering condition.
Even among the type {\bf I} operators there are examples that are not
able to violate clustering (like e.g.
$( \overline{\psi}^{(a)} P_+ \psi^{(a)} )^N$ which does not contain
a $SU(N)_L\times SU(N)_R$ singlet part). The operators with
the structure $\prod^n {\cal O}_+ \prod^m {\cal O}_- \prod J_\mu^{(l)} \;, \;
n-m \neq 0$ we call `type {\bf V}' for violating.

We have to distinguish the following
cases
 \\
\begin{center}
\begin{tabular}{c|l|l|l}
case $\#$ &
$Q_5({\scriptstyle {\cal A}})$ &
$Q_5({\scriptstyle {\cal B}})$ &
$Q_5({\scriptstyle {\cal AB}})$ \\ \hline
{\bf 1} & {\bf II} & {\bf II} & {\bf II} \\
{\bf 2} & {\bf II} $\;\;(\neq 0)$ & {\bf II}
$\;\;(\neq 0)$ & {\bf I} \\
{\bf 3} & {\bf II} $\;\;(\neq 0)$ & {\bf I} & {\bf II} \\
{\bf 4} & {\bf II} $\;\;( = 0)$ & {\bf I} & {\bf I} \\
{\bf 5} & {\bf I} $\;\;\;( = q)$ & {\bf I} $\;\;\;( = -q)$ &
{\bf II} $\;\;( = 0)$ \\
{\bf 6} & {\bf I} & {\bf I} & {\bf I} \\
\end{tabular}
\end{center}
In brackets $(\cdot)$ we denoted facts that necessarily follow.
For example in the case 2 the requirement
$Q_5({\scriptstyle {\cal A}}) \in \mbox{{\bf II}},
Q_5({\scriptstyle {\cal B}}) \in \mbox{{\bf II}},
Q_5({\scriptstyle {\cal AB}}) \in \mbox{{\bf I}}$ implies
$Q_5({\scriptstyle {\cal A}}) \neq 0$ and
$Q_5({\scriptstyle {\cal B}}) \neq 0$. If one of these two had
charge zero,
the other operator would be of type {\bf I},
since $Q_5({\scriptstyle {\cal AB}}) =
Q_5({\scriptstyle {\cal A}}) + Q_5({\scriptstyle {\cal B}})$. \\
 \\
{\bf Case 1:}
\begin{equation}
C_\theta(\tau) = \langle {\cal A} (\tau) {\cal B} \rangle -
\langle {\cal A } \rangle \langle {\cal B} \rangle
\stackrel{\tau \rightarrow \infty}{\longrightarrow} 0 \; ,
\end{equation}
since if $\cal A$ and $\cal B$ are of type {\bf II}, they do not
form a pair that violates clustering. \\
{\bf Case 2:}
\[
C_\theta(\tau) = \lim_{\tau^\prime \rightarrow \infty}
\langle {\cal U}_{\tau^\prime} ( {\scriptstyle {\cal AB}})
{\cal A}(\tau) {\cal B} \rangle \; - \; \langle {\cal A} \rangle
\langle {\cal B} \rangle \stackrel{\tau \rightarrow \infty}{\longrightarrow} \]
\begin{equation}
\langle {\cal A} \rangle \lim_{\tau^\prime \rightarrow \infty}
\langle {\cal U}_{\tau^\prime} ( {\scriptstyle {\cal AB}})
{\cal B} \rangle \; - \; \langle {\cal A} \rangle
\langle {\cal B} \rangle \; = \; 0 \; ,
\end{equation}
where we used the fact that $\langle {\cal A} \rangle$ factorizes, since
$\cal A$ is type {\bf II} and $\langle {\cal A} \rangle = 0$ since
$Q_5({\scriptstyle {\cal A}}) \neq 0$ (compare the note in the table).
The interchange of the $\tau , \tau^\prime$ limits is justified, since
all the functions involved are continuous in these variables and bounded
(the exponent $E$ in equation (107) is nonnegative). \\
{\bf Case 3:}
\begin{equation}
C_\theta(\tau) = \langle {\cal A}(\tau) {\cal B} \rangle \; - \;
\langle {\cal A} \rangle \lim_{\tau^\prime \rightarrow \infty}
\langle {\cal U}_{\tau^\prime} ( {\scriptstyle {\cal B}})
{\cal B} \rangle \stackrel{\tau \rightarrow \infty}{\longrightarrow} 0 \; ,
\end{equation}
for the same reasons as in the last case. \\
{\bf Case 4:}
\[
C_\theta(\tau) = \lim_{\tau^\prime \rightarrow \infty}
\langle {\cal U}_{\tau^\prime} ( {\scriptstyle {\cal AB}})
{\cal A}(\tau) {\cal B} \rangle \; - \; \langle {\cal A} \rangle
\lim_{\tau^\prime \rightarrow \infty}
\langle {\cal U}_{\tau^\prime} ( {\scriptstyle {\cal B}})
{\cal B} \rangle \stackrel{\tau \rightarrow \infty}{\longrightarrow}  \]
\begin{equation}
\langle {\cal A} \rangle  \lim_{\tau^\prime \rightarrow \infty}
\langle {\cal U}_{\tau^\prime} ( {\scriptstyle {\cal AB}})
{\cal B} \rangle \; - \; \langle {\cal A} \rangle
\lim_{\tau^\prime \rightarrow \infty}
\langle {\cal U}_{\tau^\prime} ( {\scriptstyle {\cal B}})
{\cal B} \rangle = 0 \; .
\end{equation}
Again $\langle {\cal A} \rangle$ factorizes, and
${\cal U}_{\tau^\prime} ( {\scriptstyle {\cal AB}}) =
{\cal U}_{\tau^\prime} ( {\scriptstyle {\cal B}})$ since
$Q_5({\scriptstyle {\cal AB}}) = Q_5({\scriptstyle {\cal B}})$. \\
{\bf Case 5:}
\[
C_\theta(\tau) = \langle {\cal A}(\tau) {\cal B} \rangle \; - \;
\lim_{\tau^\prime \rightarrow \infty} \langle {\cal U}_{\tau^\prime}
( {\scriptstyle {\cal A}}) {\cal A} \rangle
\lim_{\tau^{\prime \prime} \rightarrow \infty} \langle
{\cal U}_{\tau^{\prime \prime}} ( {\scriptstyle {\cal B}}) {\cal B} \rangle
\stackrel{\tau \rightarrow \infty}{\longrightarrow} \]
\begin{equation}
\left\{ \begin{array}{l}
{\cal F_A  F_B - F_A  F_B}  = 0 \; \; \mbox{for} \; \; {\cal A,B} \; \;
\mbox{of {\bf V}-type} \; , \\
0 \; \; \mbox{otherwise} \; . \end{array} \right.
\end{equation}
In the first case we used the factorization of the argument function
$\cal F_{AB}$ introduced in (118). \\
{\bf Case 6:}
\[
C_\theta(\tau) = \lim_{\tau^\prime \rightarrow \infty}
\langle {\cal U}_{\tau^\prime} ( {\scriptstyle {\cal AB}})
{\cal A}(\tau) {\cal B} \rangle \; - \;
\lim_{\tau^\prime \rightarrow \infty} \langle {\cal U}_{\tau^\prime}
( {\scriptstyle {\cal A}}) {\cal A} \rangle
\lim_{\tau^{\prime \prime} \rightarrow \infty} \langle
{\cal U}_{\tau^{\prime \prime}} ( {\scriptstyle {\cal B}}) {\cal B} \rangle
\stackrel{\tau \rightarrow \infty}{\longrightarrow} \]
\begin{equation}
\left\{ \begin{array}{l}
{\cal F_A} \lim_{\tau^\prime \rightarrow \infty}
{\cal F}_{{\cal U}_{\tau^\prime} {\cal B}} -
{\cal F_A  F_B}  = 0 \; \; \mbox{for} \; \; {\cal A,B} \; \;
\mbox{of {\bf V}-type} \; , \\
0 \; \; \mbox{otherwise} \; . \end{array} \right.
\end{equation}
To justify the first case, one has to show that
$\lim_{\tau^\prime \rightarrow \infty}
{\cal F}_{{\cal U}_{\tau^\prime} {\cal B}} = {\cal F_B}$. This can be
seen immediately from (119). Shifting $\tau^\prime$ in
${\cal F}_{{\cal U}_{\tau^\prime} {\cal B}}$ corresponds to shifting the
arguments of the test operator
${\cal U}_{\tau^\prime}( {\scriptstyle {\cal B}})$. For
example this could be
the set (refering to (119) for $\cal F$)
\begin{equation}
\left\{ x_i^{(a)} \mid n^\prime < i \leq n ; a = 1,2,...N \right\} \; ,
\end{equation}
where $n^\prime < n$ depends on the charge $Q_5( {\scriptstyle {\cal B}})$.
The $\tau^\prime$ terms cancel for $\tau^\prime \rightarrow \infty$,
and what remains is ${\cal F_B}$, since the normalization of
${\cal U}_{\tau^\prime}( {\scriptstyle {\cal B}})$ cancels exactly the
extra terms.
$\Box$

This concludes our analysis of the vacuum structure of the $N$ flavor
Schwinger model. The result is in full agreement with the picture that
is conventionally deduced from the discussions of topological sectors
\cite{joos},\cite{wipf}.


\end{document}